\documentclass[a4paper,fleqn,usenatbib,]{mnras}

\hypersetup{pdfauthor={T. Troester},
               			     bookmarksnumbered=true}
			     
\usepackage{graphicx}	
\usepackage{amsmath}	
\usepackage{amssymb}	

\usepackage{braket}
\usepackage{xspace}

\usepackage[caption=false]{subfig}

\usepackage{environ}
\NewEnviron{splitequation}{%
	\begin{equation}\begin{split}
		\BODY
	\end{split}\end{equation}}

\renewcommand{\vec}{\boldsymbol}
\newcommand{\mat}{\mathbfss}

\newcommand{\highlight}[1]{}
\newcommand{\software}{\textsc}

\newcommand{\mr}[1]{\mathrm{#1}}

\newcommand{\transpose}{^\mr{T}}

\newcommand{\e}{\mathrm{e}}
\newcommand{\diff}{\mathrm{d}}

\newcommand{\units}[1]{\ \mathrm{#1}}

\newcommand{\ee}[1]{\times10^{#1}}

\newcommand{\DM}{dark matter\xspace}

\newcommand{\de}{\mathrm d}

\newcommand{\ave}{{\sc mid}}
\newcommand{\low}{{\sc low}}
\newcommand{\high}{{\sc high}}
\newcommand{\sv}{\langle\sigma_\mr{ann} v \rangle}
\newcommand{\mdm}{{m_{\rm DM}}}
\newcommand{\odm}{{\Omega_\mathrm{DM}}}
\newcommand{\om}{{\Omega_\mr{M}}}

\newcommand{\Fermi}{\textit{Fermi-LAT}\xspace}

\title[Cross-correlation of lensing and gamma rays]{Cross-correlation of weak lensing and gamma rays: implications for the nature of dark matter}
\author[T. Tr\"oster et al.]{Tilman Tr\"oster,$^{1}$\thanks{E-mail: troester@phas.ubc.ca}
Stefano Camera,$^{2}$
Mattia Fornasa,$^{3}$
Marco Regis,$^{4,5}$\newauthor
Ludovic van Waerbeke,$^{1}$
Joachim Harnois-D\'eraps,$^{6}$
Shin'ichiro Ando,$^{3}$\newauthor
Maciej Bilicki,$^{7}$
Thomas Erben,$^{8}$
Nicolao Fornengo,$^{4,5}$
Catherine Heymans,$^{6}$\newauthor
Hendrik Hildebrandt,$^{8}$
Henk Hoekstra,$^{7}$
Konrad Kuijken$^{7}$
and Massimo Viola$^{7}$
\\
$^{1}$Department of Physics and Astronomy, The University of British Columbia, 6224 Agricultural Road, Vancouver, B.C., V6T 1Z1, Canada\\
$^{2}$Jodrell Bank Centre for Astrophysics, The University of Manchester, Alan Turing Building, Oxford Road, Manchester M13 9PL, UK\\
$^{3}$Gravitation Astroparticle Physics Amsterdam (GRAPPA), University of Amsterdam, Science Park 904, 1098 XH Amsterdam, Netherlands\\
$^{4}$Dipartimento di Fisica, Universit\`{a} di Torino, via P. Giuria 1, I--10125 Torino, Italy\\
$^{5}$Istituto Nazionale di Fisica Nucleare, Sezione di Torino, via P. Giuria 1, I--10125 Torino, Italy\\
$^{6}$Institute for Astronomy, University of Edinburgh, Royal Observatory, Blackford Hill, Edinburgh, EH9 3HJ, UK\\
$^{7}$Leiden Observatory, Leiden University, P.O. Box 9513 2300 RA Leiden, The Netherlands\\
$^{8}$Argelander-Institut f\"ur Astronomie, Auf dem H\"ugel 71, 53121 Bonn, Germany\\
}

\date{Accepted 2017 February 8. Received 2017 February 3; in original form 2016 November 23}

\pubyear{2017}

\begin{document}
\label{firstpage}
\pagerange{\pageref{firstpage}--\pageref{lastpage}}
\maketitle

\begin{abstract}
We measure the cross-correlation between \Fermi gamma-ray photons and over 1000 deg$^2$ of weak lensing data from the Canada-France-Hawaii Telescope Lensing Survey (CFHTLenS), the Red Cluster Sequence Lensing Survey (RCSLenS), and the Kilo Degree Survey (KiDS). We present the first measurement of tomographic weak lensing cross-correlations and the first application of spectral binning to cross-correlations between gamma rays and weak lensing.
The measurements are performed using an angular power spectrum estimator while the covariance is estimated using an analytical prescription. We verify the accuracy of our covariance estimate by comparing it to two internal covariance estimators. 
Based on the non-detection of a cross-correlation signal, we derive constraints on weakly interacting massive particle (WIMP) dark matter. We compute exclusion limits on the dark matter annihilation cross-section $\sv$, decay rate $\Gamma_\mr{dec}$, and particle mass $\mdm$. We find that in the absence of a cross-correlation signal, tomography does not significantly improve the constraining power of the analysis. Assuming a strong contribution to the gamma-ray flux due to  small-scale clustering of dark matter and accounting for known astrophysical sources of gamma rays, we exclude the thermal relic cross-section for particle masses of $\mdm\la 20$~GeV. 
\end{abstract}


\section{Introduction}
The matter content of the Universe is dominated by so called dark matter whose cosmological abundance and large scale clustering properties have been measured to high precision \citep[e.g.][]{WMAP9, Planck2015, Ross2015, Anderson2014, Mantz2015, Hoekstra2015, Hildebrandt2017}. 
However, little is known about its microscopic nature, beyond its lack of -- or at most weak -- non-gravitational interaction with standard model matter.

Weakly interacting massive particles (WIMPs) thermally produced in the early Universe are among the leading \DM candidates. With a mass of the order of GeV/TeV, their decoupling from thermal equilibrium occurs in the non-relativistic regime. 
The weak interaction rate with lighter standard model particles furthermore ensures that their thermal relic density is naturally of the order of the measured cosmological \DM abundance \citep{Lee1977,Gunn1978}.

Many extensions of the standard model of particle physics predict the existence of new massive particles at the weak scale; some of these extra states can indeed be `dark', i.e., be colour and electromagnetic neutral, with the weak force and gravity as the only relevant coupling to ordinary matter \citep[for reviews, see e.g. ][]{Jungman1996,Bertone2005,Schmaltz2005,Hooper2007,Feng2010}.

The weak coupling allows us to test the hypothesis of WIMP \DM: supposing that WIMPs are indeed the building blocks of large scale structure (LSS) in the Universe, there is a small but finite probability that WIMPs in \DM haloes annihilate or decay into detectable particles.
These standard model particles produced by these annihilations or decays would manifest as cosmic rays which can be observed. In particular, since the WIMP mass is around the electroweak scale, gamma rays can be produced, which can be observed with ground-based or space-borne telescopes, e.g., the \Fermi telescope \citep{Atwood2009}. Indeed, analyses of the gamma-ray sky have already been widely used to put constraints on WIMP \DM, see e.g. \cite{Charles2016} for a recent review. 

The currently strongest constraints on the annihilation cross-section and WIMP mass come from the analysis of local regions with high \DM content, such as dwarf spheroidal galaxies (dSphs) \citep{Ackermann2015b}. These analyses exclude annihilation cross-sections larger than $\sim 3\ee{-26}\units{cm^3 s^{-1}}$ for \DM candidates lighter than 100 GeV. This value for the annihilation cross-section is known as the thermal cross-section and, below it, many models of new physics predict \DM candidates that yield a relic \DM density in agreement with cosmological measurements of the \DM abundance \citep{Jungman1996}.

Instead of these local probes of \DM properties, one could consider the unresolved gamma-ray background (UGRB), i.e., the cumulative radiation produced by all sources that are not bright enough to be resolved individually. Correctly modelling the contribution of astrophysical sources, such as blazars, star-forming, and radio galaxies, allows the measurement of the UGRB to be used to constrain the component associated with \DM \citep{Fornasa2015}. Indeed, the study of the energy spectrum of the UGRB \citep{The-Fermi-LAT-Collaboration2015a}, as well as of its anisotropies \citep{Ando2013,Fornasa2016} and correlation with tracers of LSS \citep{Ando2014,Shirasaki2014, Fornengo2015, Regis2015, Cuoco2015, Shirasaki2015, Ando2016,Shirasaki2016,Feng2017} have yielded independent and competitive constraints on the nature of \DM.

In this paper, we focus on the cross-correlation of the UGRB with weak gravitational lensing. Gravitational lensing is an unbiased tracer of matter and thus closely probes the distribution of \DM in the Universe. This makes it an ideal probe to cross-correlate with gamma rays to investigate the particle nature of \DM \citep{Camera2013}.

We extend previous analyses of cross-correlations of gamma rays and weak lensing of \cite{Shirasaki2014, Shirasaki2016} by adding weak lensing data from the Kilo Degree Survey (KiDS) \citep{de-Jong2013a, Kuijken2015} and making use of the spectral and tomographic information contained within the data sets. This paper presents the first tomographic weak lensing cross-correlation measurement and the first application of spectral binning to the cross-correlation between gamma rays and galaxy lensing. Exploiting tomography and the information contained in the energy spectrum of the \DM annihilation signal has been shown to greatly increase the constraining power compared to the case where no binning in redshift or energy is performed \citep{Camera2015}. 

The structure of this paper is as follows: in Section \ref{sec:models} we introduce the formalism and theory; the data sets are described in Section \ref{sec:data}; Section \ref{sec:methods} introduces the measurement methods and estimators; the results are presented in Section \ref{sec:results}; and we draw our conclusions in Section \ref{sec:conclusions}.

\section{Formalism}
\label{sec:models}

Our theoretical predictions are obtained by computing the angular cross power spectrum $C_\ell^{g\kappa}$ between the lensing convergence $\kappa$ and gamma-ray emissions for different classes of gamma-ray sources, denoted by $g$. In the Limber approximation \citep{Limber1953}, it takes the form:

\begin{splitequation}
	C_\ell^{g\kappa} =& \int_{\Delta E} \de E\, \int_0^\infty \de z\,\frac{c}{H(z)}\frac{1}{\chi(z)^2} \\
				   &\times W_{g}(E,z)\, W_{\kappa}(z) \,P_{g\delta}\left(k=\frac{\ell}{\chi(z)},z\right)\;, 
	\label{eq:clgen}
\end{splitequation}
where $z$ is the redshift, $E$ is the gamma-ray energy and $\Delta E$ the energy bin that is being integrated over, $c$ is the speed of light in the vacuum, $H(z)$ is the Hubble rate, and $\chi(z)$ is the comoving distance.
We employ a flat $\Lambda$CDM cosmological model with parameters taken from \cite{Planck2015}.

$W_{g}$ and $W_{\kappa}$ are the window functions that characterize the redshift and energy dependence of the gamma-ray emitters and the efficiency of gravitational lensing, respectively.
$P_{g\delta}(k,z)$ is the three-dimensional cross power spectrum between the gamma-ray emission for a gamma-ray source class and the matter density $\delta$, with $k$ being the modulus of the wavenumber and $\ell$ the angular multipole. 
The functional form of the window functions and power spectra depend on the populations of gamma-ray emitters and source galaxy distributions under consideration and are described in the following subsections.

The quantity measured from the data is the tangential shear cross-correlation function $\xi^{g\gamma_t}(\vartheta)$. This correlation function is related to the angular cross power spectrum by a Hankel transformation:
\begin{splitequation}
	\label{equ:Cl2xi}
	\xi^{g \gamma_t}(\vartheta) = \frac{1}{2\pi} \int_0^\infty \diff\ell\ \ell J_2(\ell\vartheta) C_\ell^{g\kappa} \ ,
\end{splitequation}
where $\vartheta$ is the angular separation in the sky and $J_2$ is the Bessel function of the first kind of order two.

\subsection{Window functions}
\label{sec:wf}
The window function describes the distribution of the signal along the line of sight, averaged over all lines of sight. 
\subsubsection{Gravitational lensing}
\begin{figure}
    \begin{center}
	\includegraphics[width=\columnwidth]{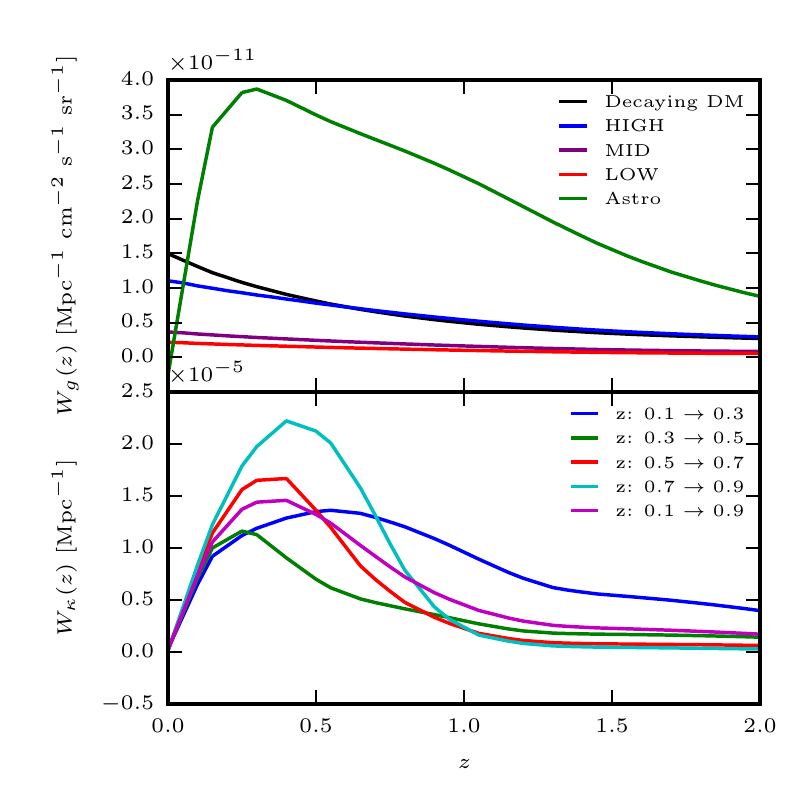}
  	\caption{\emph{Top:} window functions for the gamma-ray emissions $W_g$ for the energy range 0.5 -- 500 GeV and redshift selection of 0.1 -- 0.9. Shown are the window functions for the three annihilating \DM scenarios considered, i.e., \high\ (blue), \ave\ (purple), \low\ (red); decaying \DM (black); and the sum of the astrophysical sources (green). The annihilation scenarios assume $\mdm=100$ GeV and $\sv=3\ee{-26}\units{cm^3 s^{-1}}$. For decaying \DM, $\mdm=200$ GeV and $\Gamma_\mr{dec}=5\ee{-28}\units{s^{-1}}$. The predictions for annihilating and decaying \DM are for the $b\bar b$ channel. We consider three populations of astrophysical sources that contribute to the UGRB: blazars, mAGNs, and SFGs, described in Section~\ref{sec:W_astro}.
	\emph{Bottom:} the lensing window functions for the five tomographic bins chosen for KiDS.}
  	\label{fig:windowfunctions}
   \end{center}
\end{figure}

For the gravitational lensing the window function is given by (see, e.g., \citealt{Bartelmann2010a})
\begin{splitequation}
	W_{\kappa}(z) = \frac{3}{2} H_0^2 \Omega_\mr{M} (1+z)\chi(z) \int_{z}^\infty \de z' \, \frac{\chi(z')-\chi(z)}{\chi(z')} n(z') \ ,
	\label{eq:W_lens}
\end{splitequation}
where $H_0$ is the Hubble rate today, $\om$ is the current matter abundance in the Universe, and $n(z)$ is the redshift distribution of background galaxies in the lensing data set. The galaxy distribution depends on the data set and redshift selection, as described in Section~\ref{sec:lensing-data}. The redshift distribution function $n(z)$ is binned in redshift bins of width $\Delta_z=0.05$. To compute the window function in Eq.~\eqref{eq:W_lens}, $n(z)$ is interpolated linearly between those bins. The resulting window functions for KiDS are shown in the bottom panel of Fig.~\ref{fig:windowfunctions}. The width of the window function in Fig.~\ref{fig:windowfunctions}, especially for the 0.1--0.3 redshift bin, is due to the leakage of the photometric redshift distribution outside of the redshift selection range.

\subsubsection{Gamma-ray emission from \DM}
We consider two processes by which \DM can create gamma rays: annihilation and decay. 

The window function for annihilating \DM reads \citep{Ando2006,Fornengo2014}
\begin{splitequation}
	W_{g_\mr{ann}}(z,E) =& \frac{(\odm \rho_c)^2}{4\pi} \frac{\sv}{2\mdm^2} \left(1+z\right)^3 \Delta^2(z) \\
					&\times \frac{\de N_\mr{ann}}{\de E}\left[E(1+z) \right] \e^{-\tau\left[z,E(1+z)\right]} \ ,
	\label{eq:win_annDM}
\end{splitequation}
where $\odm$ is the cosmological abundance of \DM, $\rho_c$ is the current critical density of the Universe, $\mdm$ is the rest mass of the \DM particle, and $\sv$ denotes the velocity-averaged annihilation cross-section, assumed here to be the same in all haloes. 
$\de N_\mr{ann} / \de E$ indicates the number of photons produced per annihilation as a function of photon energy, and sets the gamma-ray energy spectrum. 
We will consider it to be given by the sum of two contributions: prompt gamma-ray production from \DM annihilation, which provides the bulk of the emission at low masses, and inverse Compton scattering of highly energetic \DM-produced electrons and positrons on CMB photons, which upscatter in the gamma-ray band. The final states of \DM annihilation are computed by means of the \software{PYTHIA} Monte Carlo package v8.160 \citep{PYTHIA2008}. The inverse Compton scattering contribution is calculated as in \cite{Fornasa2013}, which assumes negligible magnetic field and no spatial diffusion for the produced electrons and positrons.
Results will be shown for three final states of the annihilation: $b\bar b$ pairs, which yields a relatively soft spectrum of photons and electrons, mostly associated with hadronisation into pions and their subsequent decay; $\mu^+\mu^-$, which provides a relatively hard spectrum, mostly associated with final state radiation of photons and direct decay of the muons into electrons; and $\tau^+\tau^-$, which is in between the first two cases, being a leptonic final state but with semi-hadronic decay into pions \citep{Fornengo2004,Cembranos2011,Cirelli2011}.

The optical depth $\tau$ in Eq.~\eqref{eq:win_annDM} accounts for attenuation of gamma rays due to scattering off the extragalactic background light (EBL), and is taken from \cite{Franceschini2008}. 

\begin{figure}
    \begin{center}
	\includegraphics[width=\columnwidth]{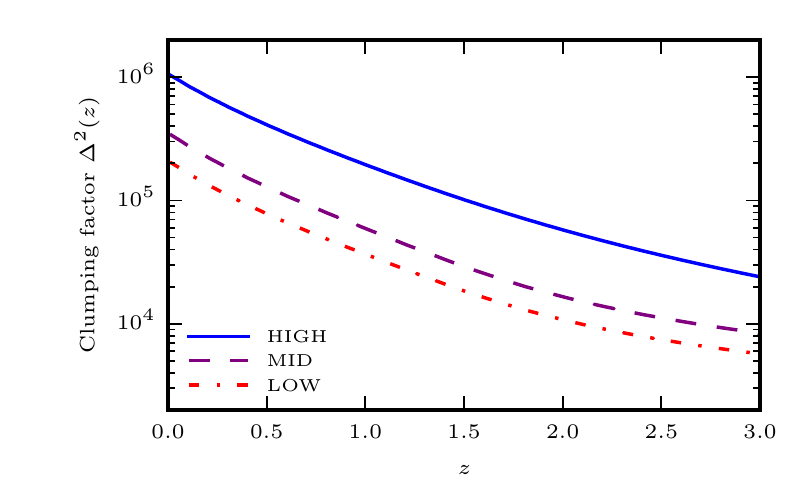}
  	\caption{Dark matter clumping factor $\Delta^2$, as defined in Eq.~\eqref{eq:clumping}, as a function of redshift for the \low\ (dash-dotted red), \ave\ (dashed purple) and \high\ (solid blue) scenarios. The \ave\ model is built from its expression at $z=0$ in \protect\cite{Moline2017}, assuming the same redshift scaling as in \protect\cite{Prada2012}.}
  	\label{fig:clumping}
   \end{center}
\end{figure}

The clumping factor $\Delta^2$ is related to how \DM density is clustered in haloes and subhaloes. Its definition depends on the square of the \DM density; therefore, it is a measure of the amount of annihilations happening and, thus, the expected gamma-ray flux.
The clumping factor is defined as \citep[see, e.g., ][]{Fornengo2014}
\begin{splitequation}
\label{eq:clumping}
	\Delta^2(z) \equiv 
	\frac{\langle \rho^2_{\rm DM} \rangle}{{\bar \rho}^2_{\rm DM}} =&
	\int_{M_{\rm min}}^{M_{\rm max}} \de M \frac{\de n_\mr{h}}{\de M}(M,z) \,\left[1+b_{\rm sub}(M,z)\right]\\ 
	&\times \int \de^3 \vec{x} \frac{\rho^2_\mr{h}({\vec{x}|M,z)}}{{\bar \rho}^2_{\rm DM}} \ ,
\end{splitequation}
where $\bar\rho_\mr{DM}$ is the current mean \DM density, $\de n_\mr{h}/\de M$ is the halo mass function~\citep{Sheth1999}, $M_{\rm min}$ is the minimal halo mass (taken to be $10^{-6} M_\odot$), $M_{\rm max}$ is the maximal mass of haloes (for definiteness, we use $10^{18} M_\odot$, but the results are insensitive to the precise value assumed), $\rho_\mr{h}(\mathbf{x}|M,z)$ is the \DM density profile of a halo with mass $M$ at redshift $z$, taken to follow a Navarro-Frenk-White (NFW) profile \citep{Navarro1997}, and $b_{\rm sub}$ is the boost factor that encodes the `boost' to the halo emission provided by subhaloes. 
To characterize the halo profile and the subhalo contribution, we need to specify their mass concentration. Modelling the concentration parameter $c(M,z)$ at such small masses and for subhaloes is an ongoing topic of research and is the largest source of uncertainty of the models in this analysis.
We consider three cases: \low, which uses the concentration parameter derived in \cite{Prada2012} \citep[see also][]{Sanchez-Conde2014}; \high, based on \cite{Gao2012}; and \ave, following the recent analysis of \cite{Moline2017}. The last one represents our reference case with predictions that are normally intermediate between those of the \low\ and \high. The authors in \cite{Moline2017} refined the estimation of the boost factor of \cite{Sanchez-Conde2014} by modelling the dependence of the concentration of the subhaloes on their position in the host halo. Accounting for this dependence and related effects, such as tidal stripping, leads to an increase of a factor $\sim 1.7$ in the overall boost factor over the \low\ model.
Predictions for the \DM clumping factor for the three models are shown in Fig.~\ref{fig:clumping}. 

Since the number of subhaloes and, therefore, the boost factor, increases with increasing host halo mass, the integral in Eq.~\eqref{eq:clumping} is dominated group and cluster-sized haloes \citep{Ando2013}. However, in the absence of subhaloes, the clumping factor in Eq.~\eqref{eq:clumping} would strongly depend on the low-mass cutoff $M_\mr{min}$. The minimum halo mass $M_\mr{min}$ depends on the free-streaming scale of \DM, which is assumed to be in the range of $10^{-12}$ -- $10^{-3} M_\odot$ \citep{Profumo2006,Bringmann2009}. We therefore choose an intermediary mass cutoff of $M_\mr{min}=10^{-6} M_\odot$. As all our models include substructure, the dependence on $M_\mr{min}$ is at most $\mathcal{O}(1)$ \citep[see, e.g., figure S3 in][]{Regis2015}.

\begin{figure}
    \begin{center}
	\includegraphics[width=\columnwidth]{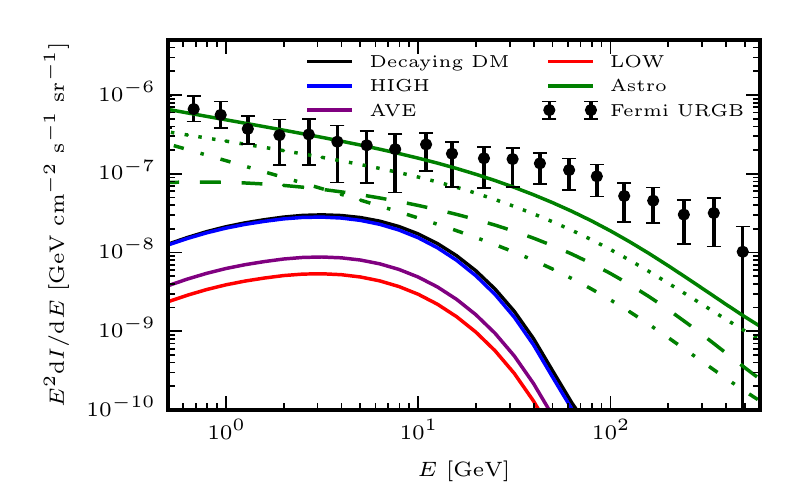}
  	\caption{Intensities of the gamma-ray source classes considered in this work: annihilating \DM assuming \high\ (solid blue), \ave\ (solid purple), \low\ (solid red) clustering models; decaying \DM (solid black); and astrophysical sources (solid green). The \DM particle properties are the same as in Fig.~\ref{fig:windowfunctions}. The astrophysical sources are further divided into blazars (dashed green), mAGN (dotted green), and SFG (dash-dotted green). The black data points represents the observed isotropic component of the UGRB \citep{Ackermann2015}.}
  	\label{fig:spectra}
   \end{center}
\end{figure}

The window function of decaying \DM is given by \citep{Ando2006,Ibarra2013,Fornengo2014}
\begin{equation}
    \label{eq:win_decDM}
    W_{g_\mr{dec}}(z,E) = \frac{\odm \rho_c}{4\pi} 
    \frac{\Gamma_{\rm dec}}{\mdm}
    \frac{\de N_\mr{dec}}{\de E} \left[E(1+z) \right] 
    \e^{-\tau\left[z,E(1+z)\right]} \ ,
\end{equation}
where $\Gamma_{\rm dec}$ is the decay rate and $\frac{\de N_\mr{dec}}{\de E}(E) = \frac{\de N_\mr{ann}}{\de E}(2E)$, i.e., the energy spectrum for decaying \DM is the same as that for annihilating \DM described above, at twice the energy \citep{Cirelli2011}. Unlike annihilating \DM, decaying \DM does not depend on the clumping factor and the expected emission is thus much less uncertain.
A set of representative window functions for annihilating and decaying \DM is shown in the top panel of Fig.~\ref{fig:windowfunctions}.
In Fig.~\ref{fig:spectra} we show the average all-sky gamma-ray emission expected from annihilating \DM for the three clumping scenarios described above and from decaying \DM. 

\subsubsection{Gamma-ray emission from astrophysical sources}
\label{sec:W_astro}
Besides \DM, gamma rays are produced by astrophysical sources which will contaminate, and even dominate over the expected \DM signal. Indeed, astrophysical sources have been shown to be able to fully explain the observed cross-correlations between gamma rays and tracers of LSS, like galaxy catalogues \citep{Xia2015,Cuoco2015}. For this analysis, we model three populations of astrophysical sources of gamma rays: blazars, misaligned active galactic nuclei (mAGNs), and star forming galaxies (SFGs). The sum of the gamma-ray emissions produced by the three extragalactic astrophysical populations described above approximately accounts for all the UGRB measured \citep[see][]{Fornasa2015}, as shown in Fig.~\ref{fig:spectra}, where the emissions from the three astrophysical source classes are compared to the most recent measurement of the UGRB energy spectrum from \cite{Ackermann2015}.
 For each of these astrophysical gamma-ray sources, we consider a window function of the form 
\begin{splitequation}
    W_{g_\mr{S}}(z,E) = \chi(z)^2 \int_{\mathcal{L}_{\rm min}}^{\mathcal{L}_{\rm max}(F_{\rm sens},z)} \de \mathcal{L} 
    \, \frac{\de F}{\de E}\left(\mathcal{L},z\right) \, \Phi(\mathcal{L},z) \ ,
    \label{eq:win_astro}
\end{splitequation}
where $\mathcal{L}$ is the gamma-ray luminosity in the energy interval 0.1 -- 100 GeV, and $\Phi$ is the gamma-ray luminosity function (GLF) corresponding to one of the source classes of astrophysical emitters included in our analysis. 
The upper bound, $\mathcal{L}_{\rm max}(F_{\rm sens},z)$, is the luminosity above which an object can be resolved, given the detector sensitivity $F_{\rm sens}$, taken from \cite{Ackermann2015a}. 
As we are interested in the contribution from unresolved astrophysical sources, only sources with luminosities smaller than $\mathcal{L}_{\rm max}$ are included.
Conversely, the minimum luminosity $\mathcal{L}_{\rm min}$ depends on the properties of the source class under investigation. 
The differential photon flux is given by $\de F/\de E=\de N_\mr{S}/\de E\times \e^{-\tau\left[z,E(1+z)\right]}$, where $\de N_\mr{S}/\de E$ is the observed energy spectrum of the specific source class and the exponential factor again accounts for the attenuation of high-energy photons by the EBL.

We consider a unified blazar model combining BL Lacertae and flat-spectrum radio quasars as a single source class.
The GLF and energy spectrum are taken from \cite{Ajello2015} where they are derived from a fit to the properties of resolved blazars in the third Fermi-LAT catalogue \citep{Acero2015}.

In the case of mAGN, we follow \cite{Di-Mauro2014a}, who studied the correlation between the gamma-ray and radio luminosity of mAGN, and derived the GLF from the radio luminosity function. We consider their best-fitting $\mathcal{L}$ -$L_{r,{\rm core}}$ relation and assume a power-law spectrum with index $\alpha_{\rm mAGN}=2.37$ . 

To build the GLF of SFG, we start from the IR luminosity function of \cite{Gruppioni2013} (adding up spiral, starburst, and SF-AGN populations of their table 8) and adopt the best-fitting $\mathcal{L}$-$L_{\rm IR}$ relation from \cite{Ackermann2012}.
The energy spectrum is taken to be a power-law with spectral index $\alpha_{\rm SFG}=2.7$.

The window function and average all-sky emission expected from the astrophysical sources are shown as green lines in the top panel of Fig.~\ref{fig:windowfunctions} and Fig.~\ref{fig:spectra}, respectively. 

\subsection{Three dimensional power spectrum}
\label{sec:ps}

The three dimensional cross-power spectrum $P_{g\delta}$ between the gamma-ray emission of a source class $g$ and the matter density is defined as
\begin{splitequation}
    \langle \hat f_{g}(z,\vec{k})\hat f_{\delta}^\ast (z^\prime,\vec{k}^\prime)\rangle=(2\pi)^3\delta^3(\vec{k}+\vec{k}^\prime)P_{g\delta}(k,z,z^\prime)\ ,
    \label{eq:PS1}
\end{splitequation}
where $\hat f_{g}$ and $\hat f_{\delta}$ denote the Fourier transform of the emission of the specific class of gamma-ray emitters and matter density, respectively, and $\langle  \, . \, \rangle$ indicates the average over the survey volume. Using the Limber approximation, one can set $z=z^\prime$ in Eq.~\ref{eq:PS1}. The density of gamma-ray emission due to decaying \DM traces the \DM density contrast $\delta_\mr{DM}$, while the emission associated with annihilating \DM traces $\delta^2_\mr{DM}$. Astrophysical sources are assumed to be point-like biased tracers of the matter distribution. Finally, lensing directly probes the matter contrast $\delta_\mr{M}$.
To compute $P_{g\delta}$, we follow the halo model formalism (for a review, see, e.g., \citealt{Cooray2002}), and write $P=P^\mr{1h}+P^\mr{2h}$.
We derive the 1-halo term $P^\mr{1h}$ and the two-halo term $P^\mr{2h}$ as in \cite{Fornengo2015} and in \cite{Camera2015}.

\subsubsection{Dark matter gamma-ray sources}
The 3D cross power spectrum between \DM sources of gamma rays and matter density is given by:
\begin{splitequation}
    P_{g_\mr{DM}\kappa}^\mr{1h}(k,z) =& \int_{M_{\rm min}}^{M_{\rm max}} \de M\ \frac{\de n_\mr{h}}{\de M}(M,z) \,\hat v_{g_\mr{DM}}(k|M,z) \, \hat u_\kappa(k|M,z) \\
    P_{g_\mr{DM}\kappa}^\mr{2h}(k,z) =& \left[\int_{M_{\rm min}}^{M_{\rm max}} \de M\,\frac{\de n_\mr{h}}{\de M}(M,z)\, b_\mr{h}(M,z) \,\hat v_{g_\mr{DM}}(k|M,z) \right]\\
                             &\times \left[\int_{M_{\rm min}}^{M_{\rm max}} \de M\,\frac{\de n_\mr{h}}{\de M}(M,z)\,b_\mr{h}(M,z) \hat u_\kappa(k|M,z) \right] \\
                             &\times  P^{\rm lin}(k,z)\ ,
	\label{eq:PSDM}
\end{splitequation}
where $P^{\rm lin}$ is the linear matter power spectrum, $b_\mr{h}$ is the linear bias (taken from the model of \citealt{Sheth1999}), and  $\hat u_\kappa(k|M,z)$ is the Fourier transform of the matter halo density profile, i.e., $\rho_\mr{h}(\vec{x}|M,z)/\bar \rho_{DM}$. The Fourier transform of the gamma-ray emission profile for \DM haloes is described by $\hat v_{g_\mr{DM}}(k|M,z)$. For decaying \DM, $\hat v_{g_\mr{DM}} = \hat u_\kappa$, i.e., the emission follows the \DM density profile. Conversely, the emission for annihilating \DM follows the square of the \DM density profile: $\hat v_{g_\mr{DM}}(k|M,z) = \hat u_\mr{ann}(k|M,z) / \Delta(z)^2$, where $\hat u_\mr{ann}$ is the Fourier transform of the square of the main halo density profile plus its substructure contribution, and $\Delta(z)^2$ is the clumping factor. The mass limits are $M_{\rm min} = 10^{-6} M_\odot$ and $M_{\rm max}=10^{18} M_\odot$ again.

\subsubsection{Astrophysical gamma-ray sources}

The cross-correlation of the convergence with astrophysical sources is sourced by the 3D power spectrum
\begin{splitequation}
     P_{g_\mr{S}\kappa}^\mr{1h}(k,z) =& \int_{\mathcal{L}_{\rm min}}^{\mathcal{L}_{\rm max}} \de \mathcal{L}\,\frac{\Phi(\mathcal{L},z)}{\langle f_\mr{S} \rangle} \frac{\de F}{\de E}\left(\mathcal{L},z\right) \hat u_\kappa(k|M(\mathcal{L},z),z) \\
     P_{g_\mr{S}\kappa}^\mr{2h}(k,z) =& \left[\int_{\mathcal{L}_{\rm min}}^{\mathcal{L}_{\rm max}} \de \mathcal{L}\, b_\mr{S}(\mathcal{L},z)\,\frac{\Phi_i(\mathcal{L},z)}{\langle f_\mr{S} \rangle} \frac{\de F}{\de E}\left(\mathcal{L},z\right) \right]\\
     &\times \left[\int_{M_{\rm min}}^{M_{\rm max}} \de M\,\frac{\de n_\mr{h}}{\de M} b_\mr{h}(M,z) \hat u_\kappa(k|M,z) \right] \\
     &\times P^{\rm lin}(k,z)\ ,
	\label{eq:PSastro}
\end{splitequation}
where $b_\mr{S}$ is the bias of gamma-ray astrophysical sources with respect to the matter density, for which we adopt $b_\mr{S}(\mathcal{L},z)=b_\mr{h}(M(\mathcal{L},z))$. That is, a source with luminosity $\mathcal{L}$ has the same bias $b_\mr{h}$ as a halo with mass $M$, with the relation $M(\mathcal{L},z)$ between the mass of the host halo $M$ and the luminosity of the hosted object $\mathcal{L}$ taken from \cite{Camera2015}. The mean flux $\langle f_\mr{S} \rangle$ is defined as $\langle f_\mr{S} \rangle=\int \de \mathcal{L}\frac{\de F}{\de E} \Phi$.

\section{Data}
\label{sec:data}
\subsection{Weak lensing data sets}
\label{sec:lensing-data}

For this study we combine CFHTLenS\footnote{http://www.cfhtlens.org/} and RCSLenS\footnote{http://www.rcslens.org/} data sets from the Canada France Hawaii Telescope (CFHT) and KiDS\footnote{http://kids.strw.leidenuniv.nl/} from the VLT Survey Telescope (VST), all of which have been optimised for weak lensing analyses. 
The same photometric redshift and shape measurement algorithms have been used in the analysis of the three surveys. However, there are slight differences in the algorithm implementation and in the shear and photometric redshift calibration, as described in the following subsections.

The sensitivity of the measurement depends inversely on the overlap area between the gamma-ray map and the lensing data, with a weaker dependence on the parameters characterizing the lensing sensitivity, i.e., the galaxy number density and ellipticity dispersion. This is due to the fact that at large scales, sampling variance dominates the contribution of lensing to the covariance and reducing the shape noise does not result in an improvement of the overall covariance. This point is further discussed in Section~\ref{sec:gaussian-cov}. 

Of the three surveys, only CFHTLenS and KiDS have full photometric redshift coverage. We choose to restrict the tomographic analysis to KiDS, as the much smaller area of CFHTLenS is expected to yield a much lower sensitivity for this measurement. 
In Section~\ref{sec:DM-interpretation} we find that tomography does not appreciably improve the exclusion limits on the \DM parameters. We thus do not lose sensitivity by restricting the tomographic analysis to KiDS in this work.

\subsubsection{CFHTLenS}
CFHTLenS spans a total area of 154 deg$^2$ from a mosaic of 171 individual MEGACAM pointings, divided into four compact regions \citep{Heymans2012}. 
Details on the data reduction are described in \cite{Erben2013}. The observations in the five bands {\it ugriz} of the survey allow for the precise measurement of photometric redshifts \citep{Hildebrandt2012}. The shape measurement with \software{lensfit} is described in detail in \cite{Miller2013}. We make use of all fields in the data set as we are not affected by the systematics that lead to field rejections in the cosmic shear analyses \citep{Kilbinger2013}. We correct for the additive shear bias for each galaxy individually, while the multiplicative bias is accounted for on an ensemble basis, as described in Section~\ref{sec:estimators}.

Individual galaxies are selected based on the Bayesian photometric redshift $z_\mr{B}$ being in the range [0.2, 1.1]. The resulting redshift distribution of the selected galaxies is obtained by stacking the redshift probability distribution function of individual galaxies, weighted by the \software{lensfit} weight. As a result of the stacking of the individual redshift PDFs, the true redshift distribution leaks outside the $z_\mr{B}$ selection range. Stacking the redshift PDFs can lead to biased estimates of the true redshift distribution of the source galaxies \citep{Choi2016} but in light of the large statistical and modelling uncertainties in this analysis these biases can be safely neglected here.

\subsubsection{RCSLenS}
The RCSLenS data consist of 14 disconnected regions whose combined total area reaches 785 deg$^2$. A full survey and lensing analysis description is given in \cite{Hildebrandt2016}. RCSLenS uses the same \software{lensfit} version as CFHTLenS but with a different size prior, as galaxy shapes are measured from \textit{i}-band images in CFHTLenS, whereas RCSLenS uses the \textit{r}-band. The additive and multiplicative shear biases are accounted for in the same fashion as in CFHTLenS.

Multi-band photometric information is not available for the whole RCSLenS footprint, therefore we use the redshift distribution estimation technique described in \cite{Harnois-Deraps2016} and \cite{Hojjati2016}. Of the three magnitude cuts considered in \cite{Hojjati2016}, we choose to select the source galaxies such that  $18<\mr{mag}_\mr{r}<26$, as this selection yielded the strongest cross-correlation signal in \cite{Hojjati2016}.
This cut is close to the $18<\mr{mag}_\mr{r}<24$ in \cite{Harnois-Deraps2016} but with the faint cutoff determined by the shape measurement algorithm. 
The redshift distribution is derived from the CFHTLenS-VIPERS sample \citep{Coupon2015}, a UV and IR extension of CFHTLenS. We stack the redshift PDF in the CFHTLenS-VIPERS sample, accounting for the RCSLenS magnitude selection, \textit{r}-band completeness, and galaxy shape measurement (\software{lensfit}) weights.

\subsubsection{KiDS}
\label{sec:data-KiDS}
The third data set considered here comes from KiDS, which currently covers 450 deg$^2$ with complete {\it ugri} four band photometry in five patches. Galaxy shapes are measured in the \textit{r}-band using the new self-calibrating \software{lensfit} \citep{Fenech-Conti2016}. Cross-correlation studies such as this work are only weakly sensitive to additive biases and, being linear in the shear, are less affected by multiplicative biases than cosmic shear studies. Nonetheless, the analysis still benefits from well-calibrated shape measurements. The residual multiplicative shear bias is accounted for on an ensemble basis, as for CFHTLenS and RCSLenS. To correct for the additive bias we subtract the \software{lensfit} weighted ellipticity means in each tomographic bin. A full description of the survey and data products is given in \cite{Hildebrandt2017}. 

We select galaxies with $0.1 \leq z_\mr{B} < 0.9$ and then further split the data into four tomographic bins [0.1, 0.3], [0.3, 0.5], [0.5, 0.7], and [0.7, 0.9].
We derive the effective $n(z)$ following the DIR method introduced in \cite{Hildebrandt2017}.

\subsection{Fermi-LAT}
\label{sec:data-Fermi}
For this work we use \Fermi data gathered until 2016 September 1, spanning over eight years of observations. We use Pass 8 event reconstruction and reduce the data using \software{Fermi Science Tools} version \verb|v10r0p5|. We select \verb|FRONT+BACK| converting events (\verb|evtype=3|) between energies of 0.5 and 500 GeV. We restrict our main analysis to \verb|ultracleanveto| photons (\verb|evclass=1024|). We verify that selecting \verb|clean| photons (\verb|evclass=256|) does not change the results of the analysis. Furthermore, we apply the cuts \verb|(DATA_QUAL>0)&&(LAT_CONFIG==1)| on the data quality. We then create full sky \software{HEALPix}\footnote{http://healpix.sourceforge.net/} photon count and exposure maps with \verb|nside=1024| \citep{Gorski2005} in 20 logarithmically spaced energy bins in the range mentioned above.

The flux map used in the cross-correlation analysis is obtained by dividing the count maps by the exposure maps in each energy bin before adding them. We have confirmed that the energy spectrum of the individual flux maps follows a broken power-law with an index of $2.34\pm0.02$, consistent with that obtained in previous studies of the UGRB \citep{The-Fermi-LAT-Collaboration2015a}. 

We also create maps for four energy bins 0.5 -- 0.766 GeV, 0.766 -- 1.393 GeV, 1.393 -- 3.827 GeV, and 3.827 -- 500 GeV. The bins are chosen such that they would contain equal photon counts for a power law spectrum with index 2.5. The flux maps for the four energy bins are computed by first dividing each energy bin into three logarithmically spaced bins, creating flux maps for these fine bins, and then adding them up.

The total flux is dominated by resolved point sources and, to a lesser extend, by diffuse Galactic emissions. To probe the unresolved component of the gamma-ray sky, we mask the 500 brightest point sources in the third Fermi point source catalogue \cite{Acero2015} with circular masks with a radius of two degrees. The remaining point sources are masked with one degree circular masks. We checked that the analysis is robust with respect to other masking strategies. 
The effect of the diffuse Galactic emission is minimized by subtracting the \verb|gll_iem_v06| model. Furthermore, we employ a 20\degr\ cut in Galactic latitude. It has been shown in \cite{Shirasaki2016} that this cross-correlation analysis is robust against the choice made for the model of diffuse Galactic emission. We have confirmed that our results are not significantly affected even in the extreme scenario of not removing the diffuse Galactic emission at all. This represents an important benefit of using cross-correlations to study the UGRB over studies of the energy spectrum alone, as in \cite{The-Fermi-LAT-Collaboration2015a}. 

The robustness of these selection and cleaning choices is demonstrated in Fig.~\ref{fig:dataselection}, where the impact of the event selection, point source masks, and cleaning of the diffuse Galactic emission on the cross-correlation signal is shown. None of these choices lead to a significant change in the measured correlation signal, highlighting the attractive feature of cross-correlation analyses that uncorrelated quantities, such as Galactic emissions and extragalactic effects like lensing, do not bias the signal.

\begin{figure}
    \begin{center}
	\includegraphics[width=\columnwidth]{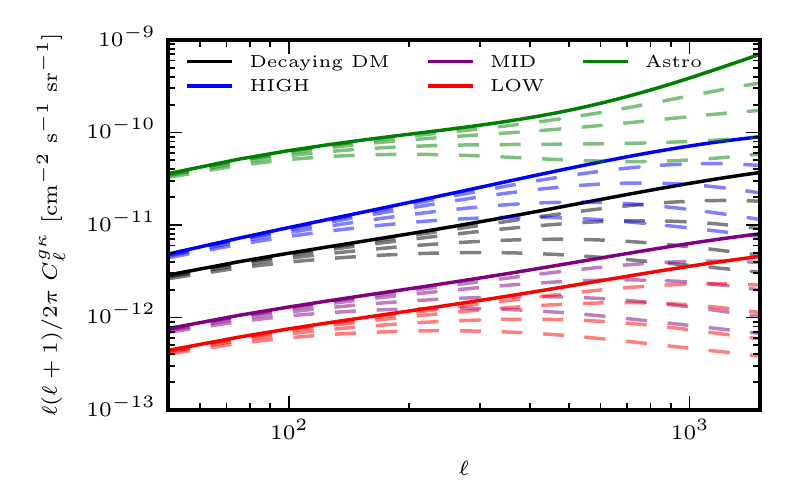}
  	\caption{Model $C_\ell^{g\kappa}$ for three annihilating \DM scenarios, decaying \DM, and astrophysical sources. The models and formatting are the same as in Figs.~\ref{fig:windowfunctions} and~\ref{fig:spectra}. The models assume the $n(z)$ for the $z\in [0.1, 0.9]$ bin for KiDS and the energy range 0.5 -- 500 GeV. The dashed lines indicate the effect of the \Fermi PSF on the cross-power spectrum for the four energy bins, with the lowest energy bin having the strongest suppression of power at small scales. For clarity, the effect of the PSF for the different energy bins is shown on the cross-power spectrum $C_\ell^{g\kappa}$ for the single energy bin of 0.5 -- 500 GeV.}
  	\label{fig:modelCl}
   \end{center}
\end{figure}

The point spread function (PSF) of \Fermi is energy dependent and, especially at low energies, significantly reduces the cross-correlation signal power at small angular scales, as demonstrated in Fig.~\ref{fig:modelCl}. The pixelation of the gamma-ray sky has a similar but much weaker effect. In this analysis, we choose to account for this suppression of power by forward modelling. That is, rather than correcting the measurements, the predicted angular power spectra $C^{g\kappa}_\ell$ are modified to account for the effect of the PSF and pixel window function.

The gamma-ray data used in the analysis are obtained by cutting out regions around the lensing footprints. To increase the sensitivity at large angular scales, we include an additional four degree wide band around each of the 23 lensing patches.

\section{Methods}
\label{sec:methods}
\subsection{Estimators}
\label{sec:estimators}
To measure the cross-correlation function between gamma rays and lensing, we employ the tangential shear estimator \cite[see also][]{Shirasaki2014, Harnois-Deraps2016, Hojjati2016}:
\begin{splitequation}
	\label{equ:xiest}
	\hat\xi^{g\gamma_{t/x}}(\vartheta) &= \frac{\sum_{ij} w_i e_{ij}^{t/x} g_j \Delta_{ij}(\vartheta)}{\sum_{ij} w_i  \Delta_{ij}(\vartheta)} \frac{1}{1+K(\vartheta)} \ , \\
	 \frac{1}{1+K(\vartheta)}  &= \frac{\sum_{ij} w_i  \Delta_{ij}(\vartheta)}{\sum_{ij} w_i(1+m_i) \Delta_{ij}(\vartheta)} \ ,
\end{splitequation}
where the sum runs over all galaxies $i$ and pixels $j$ of the gamma-ray flux map, $w_i$ is the \software{lensfit}-weight of galaxy $i$ and $e_{ij}^{t/x}$ is the tangential (t) or cross (x) component of the shear with respect to the position of pixel $j$, $g_j$ is the flux at pixel $j$, and $\Delta_{ij}(\vartheta)$ accounts for the angular binning, being equal to 1 if the distance between galaxy $i$ and pixel $j$ falls within the angular bin centred on $\vartheta$ and 0 otherwise. The factor of $\frac{1}{1+K}$ accounts for the multiplicative shear bias, with $m_i$ being the multiplicative shear bias of galaxy $i$.

The $\hat\xi^{g\gamma_{t/x}}(\vartheta)$ measurement described with Eq.~\eqref{equ:xiest} exhibits strong correlation between the angular bins at all scales. This complicates the estimation of the covariance matrix as the off-diagonal elements have to be estimated accurately. On the other hand, the covariance of the angular cross-power spectrum $\hat C^{g\kappa}_\ell$ is largely diagonal since the measurement is noise-dominated. We thus choose to work with angular power spectrum $\hat C^{g\kappa}_\ell$ instead of the correlation function $\hat\xi^{g\gamma_{t}}(\vartheta)$.
Inverting the relation in Eq.~\eqref{equ:Cl2xi}, one can construct an estimator for the angular cross power spectrum $\hat C^{g\kappa}_\ell$ based on the measurement of $\hat\xi^{g\gamma_t}(\vartheta)$ \citep{Schneider2002, Szapudi2001}. Specifically, working in the flat-sky approximation, one can write
\begin{splitequation}
	\label{equ:xi2Cl}
	\hat C^{g\kappa}_\ell = 2\pi \int_0^\infty \diff\vartheta\ \vartheta J_2(\ell\vartheta)  \hat\xi^{g\gamma_t}(\vartheta) \ .
\end{splitequation}
This estimator yields an estimate for the cross-power spectrum between the gamma rays and the E-mode of the shear field. Replacing the tangential shear $\gamma_t$ in Eq.~\eqref{equ:xi2Cl} with the cross component of the shear $\gamma_x$ results in an estimate of the cross-power spectrum between the gamma rays and the B-mode of the shear field, which is expected to vanish in the absence of lensing systematics.
In Appendix \ref{sec:Cltest}, we check that this estimator indeed accurately recovers the underlying power spectrum.

To estimate the power spectrum using estimator in Eq.~\eqref{equ:xi2Cl}, we measure the tangential shear between 1 and 301 arc minutes, in 300 linearly spaced bins. The resulting power spectrum is then binned in 5 linearly spaced bins between $\ell$ of 200 and 1500. At smaller scales the \Fermi PSF suppresses power, especially at low energies. At very large scales of $\ell\la 100$, the covariance is affected by residuals from imperfect foreground subtraction, hence we restrict ourselves to scales of $\ell > 200$.

\subsection{Covariances}
Our primary method to estimate the covariance relies on a Gaussian analytical prescription. This is justified because the covariance is dominated by photon and shape noise, both of which can be modelled accurately. To verify that this analytical prescription is a good estimate of the true covariance, we compare it to two internal covariance estimators which estimate the covariance from the data. In the first, we select random patches on the gamma-ray flux map and correlate them with the lensing data, as described in Section~\ref{sec:random-patches}. For the second method we randomise the pixels of the gamma-ray flux map within the patches used in the cross-correlation measurement, described in Section~\ref{sec:random-flux}.

Unlike an analytical covariance, inverting covariances estimated from a finite number of realizations incurs a bias \citep{Kaufman1967, Hartlap2007, Taylor2013, Sellentin2016}. The bias is dependent on the number of degrees of freedom in the measurement. Combining measurements of multiple energy or redshift bins increases the size of the measurement vector. Specifically, in the case of no binning in redshift or energy, the data vector has five elements, when binning in either redshift or energy, it contains 20 elements, and when binning in both redshift and energy, its length is 80. For a fixed number of realizations, the bias therefore changes depending on which data are used in the analysis, diminishing the advantage gained by combining multiple energy or redshift bins and making comparisons between different binning strategies harder. For this reason, we choose the analytical prescription as our primary method to estimate the covariance. 

\begin{figure*}
    \begin{center}
    \includegraphics[width=\textwidth]{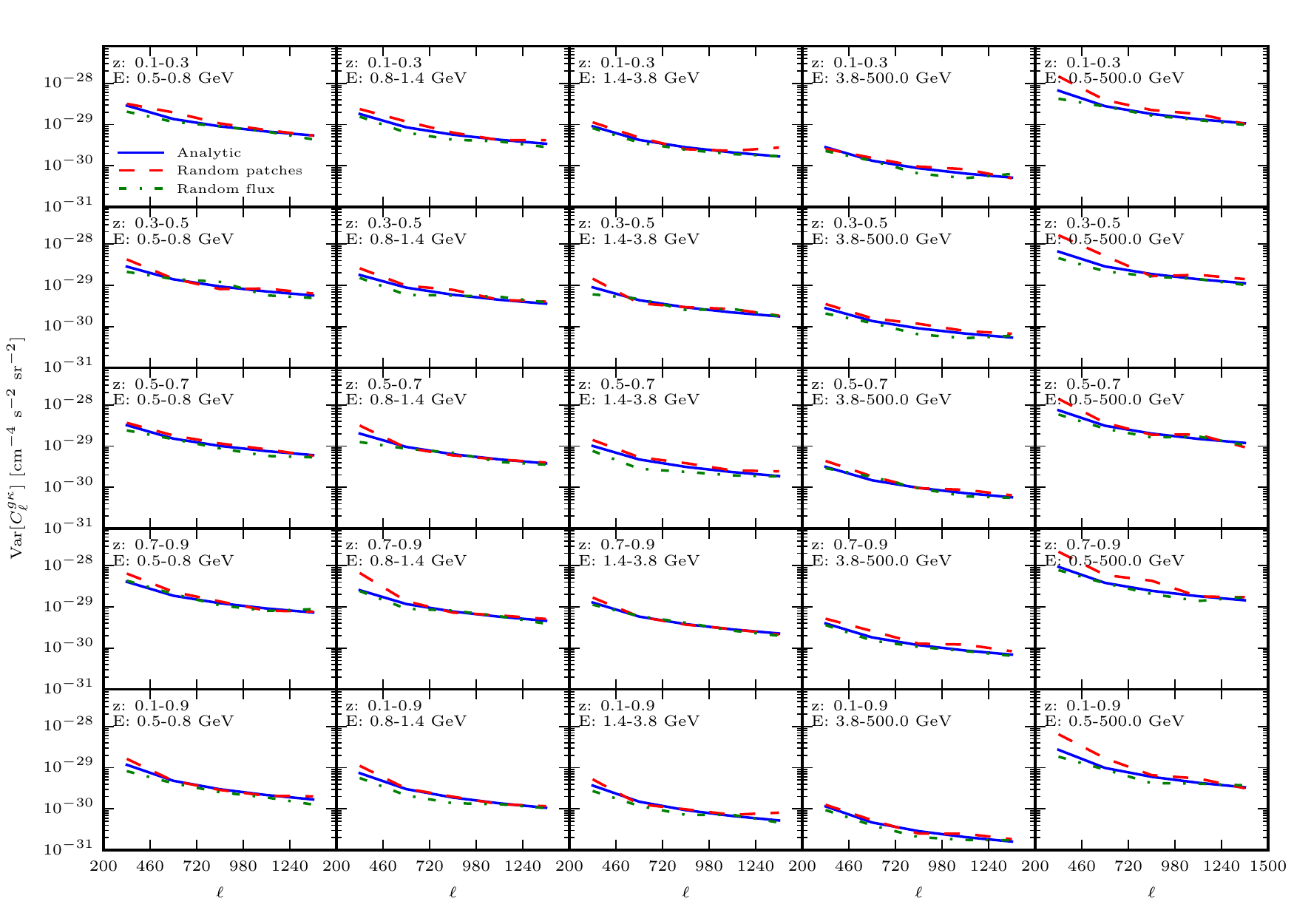}
    \caption{The diagonal elements of the analytical covariance (solid blue), covariance from random patches (dashed red), and covariance from randomized flux (dot-dashed green) for the five energy and redshift bins for KiDS. All three estimates agree at small scales, while the covariance derived from random patches shows a slight excess of variance at large scales.}
    \label{fig:KiDS-Cl-cov-5x5}
    \end{center}
\end{figure*}

The diagonal elements of the three covariance estimates are shown in Fig.~\ref{fig:KiDS-Cl-cov-5x5} for the case of KiDS, showing good agreement between all three approaches. The limits derived from the three covariance estimations agree as well. Choosing the analytical prescription as our primary method is thus justified.

\subsubsection{Analytical covariance}
\label{sec:gaussian-cov}
We model the covariance $\mat{C}$ as
\begin{splitequation}
	\label{equ:gauss-cov}
	\mat{C}[C^{g\kappa}_\ell] = \frac{1}{f_\mr{sky}(2\ell + 1)\Delta_\ell}\left( \hat C^{gg}_\ell \hat C^{\kappa\kappa}_\ell  + \left(\hat C^{g\kappa}_\ell \right)^2\right) \ ,
\end{splitequation}
where $f_\mr{sky}$ denotes the fraction of the sky that is covered by the effective area of the survey, $\Delta_\ell$ is the $\ell$-bin width, $\hat C^{gg}_\ell$ is an estimate of the gamma-ray auto-power spectrum, $\hat C^{\kappa\kappa}_\ell$ is the convergence auto-power spectrum, and $\hat C^{g\kappa}_\ell$ is the cross-spectrum between gamma rays and the convergence, calculated as described in Section~\ref{sec:models}. The effective area for the cross-correlation is given by the product of the masks of the gamma-ray map and lensing data, which corresponds to 99, 308, and 362 deg$^2$ for CFHTLenS, RCSLenS, and KiDS, respectively.

The gamma-ray auto-power spectrum $\hat C^{gg}_\ell$ is estimated from the same gamma-ray flux maps as used in the cross-correlation. We measure the auto-spectra of the five energy bins and the cross-spectra between the energy bins using \software{PolSpice}\footnote{http://www2.iap.fr/users/hivon/software/PolSpice/} in 15 logarithmically spaced $\ell$-bins between $\ell$ of 30 and 2000. Because the measurement is very noisy at large scales, we fit the measured spectra with a spectrum of the form
\begin{splitequation}
	\label{equ:Cgg-fit}
	\hat C^{gg}_\ell = C_\mr{P} + c\ \ell^\alpha \ ,
\end{splitequation}
where $C_\mr{P}$ is the Poisson noise term, and $c$ and $\alpha$ describe a power-law contribution to account for a possible increase of power at very large scales. %
The values of the intercept $c$ is consistent with zero in all cases while best-fitting Poisson noise terms are consistent with a direct estimate based on the mean number of photon counts, i.e.,
\begin{splitequation}
	C_\mr{P} = \frac{\braket{n_g/ \epsilon^2}}{\Omega_\mr{pix}} \ ,
\end{splitequation}
where $n_g$ is the number of observed photons per pixel, $\epsilon$ the exposure per pixel, and $\Omega_\mr{pix}$ the solid angle covered by each pixel \citep{Fornasa2016}. Except for the lowest energies, the observed intrinsic angular auto-power spectrum is sub-dominant to the photon shot noise \citep{Fornasa2016}.

The lensing auto-power spectrum is given by
\begin{splitequation}
	\label{equ:lensing-cov}
	\hat C^{\kappa\kappa}_\ell = C^{\kappa\kappa}_\ell + \frac{\sigma_e^2}{n_\mr{eff}} \ ,
\end{splitequation}
where $C^{\kappa\kappa}_\ell$ is the cosmic shear signal and $\frac{\sigma_e^2}{n_\mr{eff}}$ is the shape noise term, with $\sigma_e^2$ being the dispersion per ellipticity component and $n_\mr{eff}$ the galaxy number density. These parameters are listed in Table~\ref{tab:lensing-parameters}. The cosmic shear term $C^{\kappa\kappa}_\ell$ is calculated using the halo-model. The two terms in Eq.~\eqref{equ:lensing-cov} are of similar magnitude, with the shape noise dominating at small scales and sampling variance dominating at large scales. Decreasing $\sigma_\mr{e}$ or increasing $n_\mr{eff}$ thus only improves the covariance at scales where the shape noise makes a significant contribution to Eq.~\eqref{equ:lensing-cov}. However, increasing the area of the lensing survey and thus the overlap with the gamma-ray map directly decreases the covariance inversely proportionally to the overlap area. For this reason CFHTLenS has a low sensitivity in this analysis, even though it is the deepest survey of the three. Although RCSLenS has the largest effective area, the covariance for KiDS is slightly smaller since the increase in depth is large enough to overcome the area advantage of RCSLenS.

\begin{table}
    \begin{center}
         \caption{Total number of galaxies with shape measurements $n_\mr{gal}$, effective galaxy number density $n_\mr{eff}$, and ellipticity dispersion $\sigma_\mr{e}$ for CFHTLenS, RCSLenS, and KiDS for the cuts employed in this analysis. We follow the prescription in \protect\cite{Heymans2012} to calculate $n_\mr{eff}$ and $\sigma_\mr{e}$.}
	\begin{tabular}{lrcr}
\hline
    &$n_\mathrm{gal}$     &$n_\mathrm{eff}$ [arcmin$^{-2}$]   &$\sigma_\mathrm{e}$ \\
\hline
CFHTLenS   &4760606    &9.44    &0.279 \\
RCSLenS   &14490842    &5.84    &0.277 \\
KiDS $0.1\leq z_\mathrm{B} < 0.3$   &3769174    &2.23    &0.290 \\
KiDS $0.3\leq z_\mathrm{B} < 0.5$   &3249574    &2.03    &0.282 \\
KiDS $0.5\leq z_\mathrm{B} < 0.7$   &2941861    &1.81    &0.273 \\
KiDS $0.7\leq z_\mathrm{B} < 0.9$   &2640577    &1.49    &0.276 \\
KiDS $0.1\leq z_\mathrm{B} < 0.9$   &12601186    &7.54    &0.281 \\
\hline
	\end{tabular}
	\label{tab:lensing-parameters}
   \end{center}
\end{table}%

\subsubsection{Random patches}
\label{sec:random-patches}
We select 100 random patches from the gamma-ray map as an approximation of independent realizations of the gamma-ray sky. The patches match the shape of the original gamma-ray cutouts, i.e., the lensing footprints plus a four degree wide band, but have their position and orientation randomised. The patches are chosen such that they do not lie within the Galactic latitude cut.

These random patches are uncorrelated with the lensing data but preserve the auto-correlation of the gamma rays and hence account for sampling variance in the gamma-ray sky, including residuals of the foreground subtraction. 

For small patches, the assumption of independence is quite accurate, as the probability of two random patches overlapping is low. Larger patches will correlate to a certain degree. This lack of independence might lead to an underestimation of the covariance. This correlation is minimized by rotating each random patch, making the probability of having two very similar patches low.

The diagonal elements of the resulting covariance are shown in Fig.~\ref{fig:KiDS-Cl-cov-5x5}. While the agreement with the Gaussian covariance is good at small scales, there is an excess of variance at large scales for some energy and redshift bins. This excess can be explained by a large scale modulation of the power in the gamma-ray map, which would be sampled by the random patches. This interpretation is consistent with the strong growth of the error bars of the gamma-ray auto-correlation towards large scales. 
However, the results of the analysis are not affected significantly by this.

\subsubsection{Randomized flux}
\label{sec:random-flux}
In a further test of the analytical covariance in Eq.~\eqref{equ:gauss-cov}, we randomize the gamma-ray pixel positions within each patch. This preserves the one-point statistics of the flux while destroying any spatial correlations. This approach is similar to the random Poisson realizations used in \cite{Shirasaki2014} but we use the actual one-point distribution of the data itself instead of assuming a Poisson distribution.

Because the pixel values are not correlated anymore, contributions to the large scale variance due to imperfect foreground subtraction or leakage of flux from point sources outside of their masks are removed.

The covariance derived from 100 such random flux maps is in good agreement with both the analytical covariance and the covariance estimated from random patches, as shown in Fig.~\ref{fig:KiDS-Cl-cov-5x5}.

\subsection{Statistical methods}
\label{sec:stats}
The likelihood function we employ to find exclusion limits on the annihilation cross-section $\sv$ or decay rate $\Gamma_\mr{dec}$ and WIMP mass $\mdm$ is given by
\begin{splitequation}
	\label{equ:likelihood}
	\mathcal{L}(\vec\alpha | \vec d) \propto \e^{-\frac{1}{2}\chi^2(\vec d, \vec\alpha)} \ ,
\end{splitequation}
with
\begin{splitequation}
	\label{equ:likelihood-chi2}
	\chi^2(\vec d, \vec\alpha) = \left(\vec d - \vec\mu(\vec\alpha)\right)\transpose \mat{C}^{-1} \left(\vec d - \vec\mu(\vec\alpha)\right) \ ,
\end{splitequation}
where $\vec d$ denotes the data vector, $\vec\mu(\vec\alpha)$ the model vector, $\vec\alpha$ the parameters considered in the fit, i.e., either the cross-section $\sv$ and the particle mass $\mdm$ or the decay rate $\Gamma_\mr{dec}$ and $\mdm$. The amplitude of the cross-correlation signal expected from astrophysical sources is kept fixed and thus does not contribute as an extra free parameter. Finally, $\mat{C}^{-1}$ is the inverse of the data covariance.

The limits on $\sv$ and $\Gamma_\mr{dec}$ correspond to contours of the likelihood surface described by Eq.~\eqref{equ:likelihood}. Specifically, for a given confidence interval $p$, the contours are given by the set of parameters $\vec\alpha_\mr{cont.}$ for which
\begin{splitequation}
	\chi^2(\vec d, \vec \alpha_\mr{cont.}) = \chi^2(\vec d, \vec\alpha_\mr{ML}) + \Delta_{\chi^2}(p) \ ,
\end{splitequation}
where $\vec\alpha_\mr{ML}$ is the maximum likelihood estimate of the parameters $\sv$ or $\Gamma_\mr{dec}$ and $\mdm$, $\chi^2$ is given by Eq.~\eqref{equ:likelihood-chi2}, and $\Delta_{\chi^2}(p)$ corresponds to the quantile function of the $\chi^2$-distribution. For this analysis we are dealing with two degrees of freedom and require $2\sigma$ contours, hence $\Delta_{\chi^2}(0.95) = 6.18$.

This approach to estimate the exclusion limits follows recent studies, such as \cite{Shirasaki2016}. It should be noted that deriving the limits on $\sv$ or $\Gamma_\mr{dec}$ for a fixed mass $\mdm$ is also common in the literature, see e.g. \cite{Fornasa2016} for a recent example. This corresponds to calculating the quantile function $\Delta_{\chi^2}(p)$ for only one degree of freedom.

Care has to be taken when using data based covariances, such as the random patches and randomized flux, as the inverse of these covariances is biased \citep{Kaufman1967, Hartlap2007}. To account for the effect of a finite number of realizations, the Gaussian likelihood in Eq.~\eqref{equ:likelihood} should be replaced by a modified $t$-distribution \citep{Sellentin2016}. Alternatively, the effect of this bias on the uncertainties of inferred parameter can be approximately corrected \citep{Hartlap2007, Taylor2014}. In light of the large systematic uncertainties in this analysis we opt for the latter approach when using the data based covariances.

\section{Results}
\label{sec:results}
\subsection{Cross-correlation measurements}
\begin{figure*}
    \begin{center}
    \includegraphics[width=1.0\textwidth]{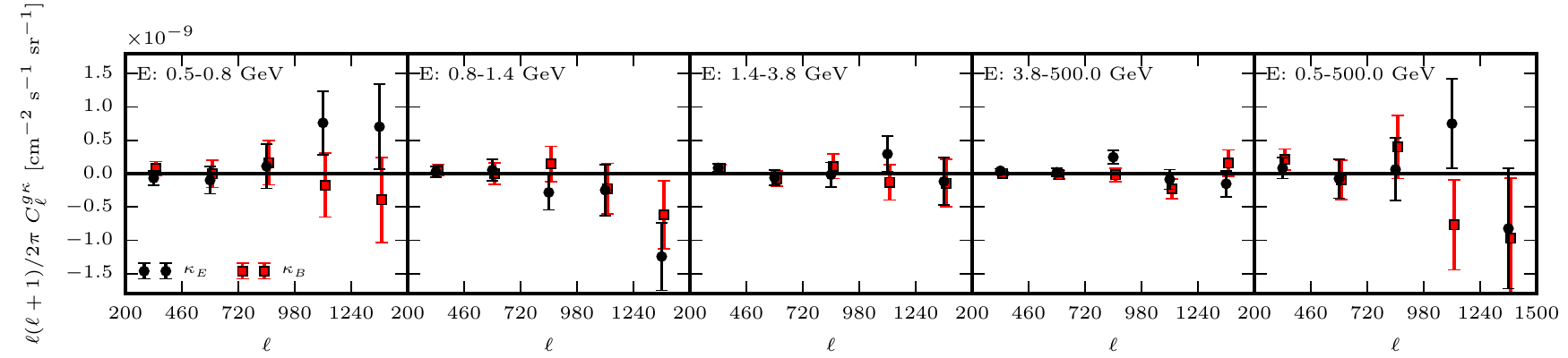}
    \caption{Measurement of the cross-spectrum $\hat C^{g\kappa}_\ell$ between \Fermi gamma rays and weak lensing data from CFHTLenS for five energy bins for gamma-ray photons (black points). The cross-spectrum of the gamma rays and CFHTLenS B-modes are depicted as red data points.}
    \label{fig:CFHTLenS-Cl-5}
    \end{center}
\end{figure*}

\begin{figure*}
    \begin{center}
    \includegraphics[width=1.0\textwidth]{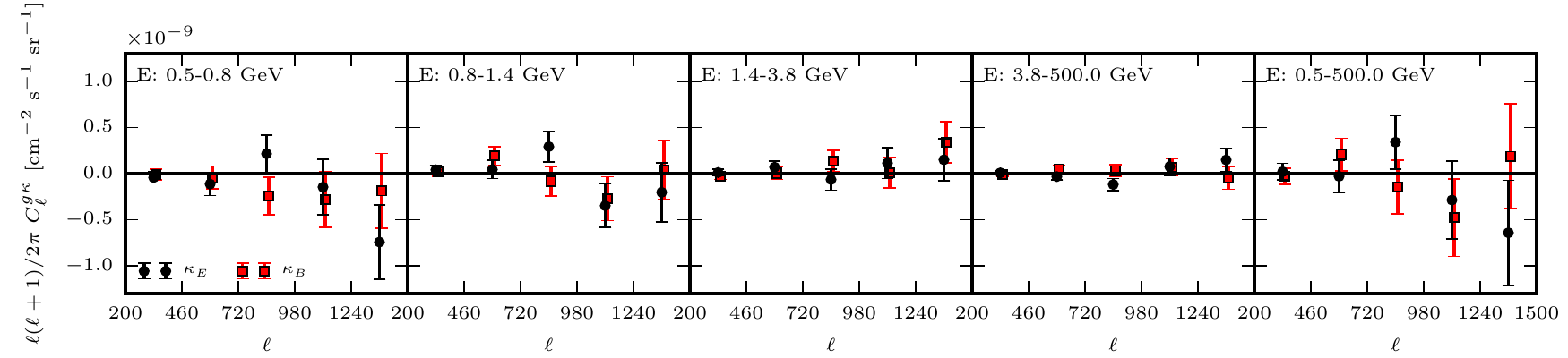}
    \caption{Measurement of the cross-spectrum $\hat C^{g\kappa}_\ell$ between \Fermi gamma rays and weak lensing data from RCSLenS for five energy bins for gamma-ray photons (black points). The cross-spectrum of the gamma rays and RCSLenS B-modes are depicted as red data points.}    
    \label{fig:RCSLenS-Cl-5}
    \end{center}
\end{figure*}

\begin{figure*}
    \begin{center}
    \includegraphics[width=1.0\textwidth]{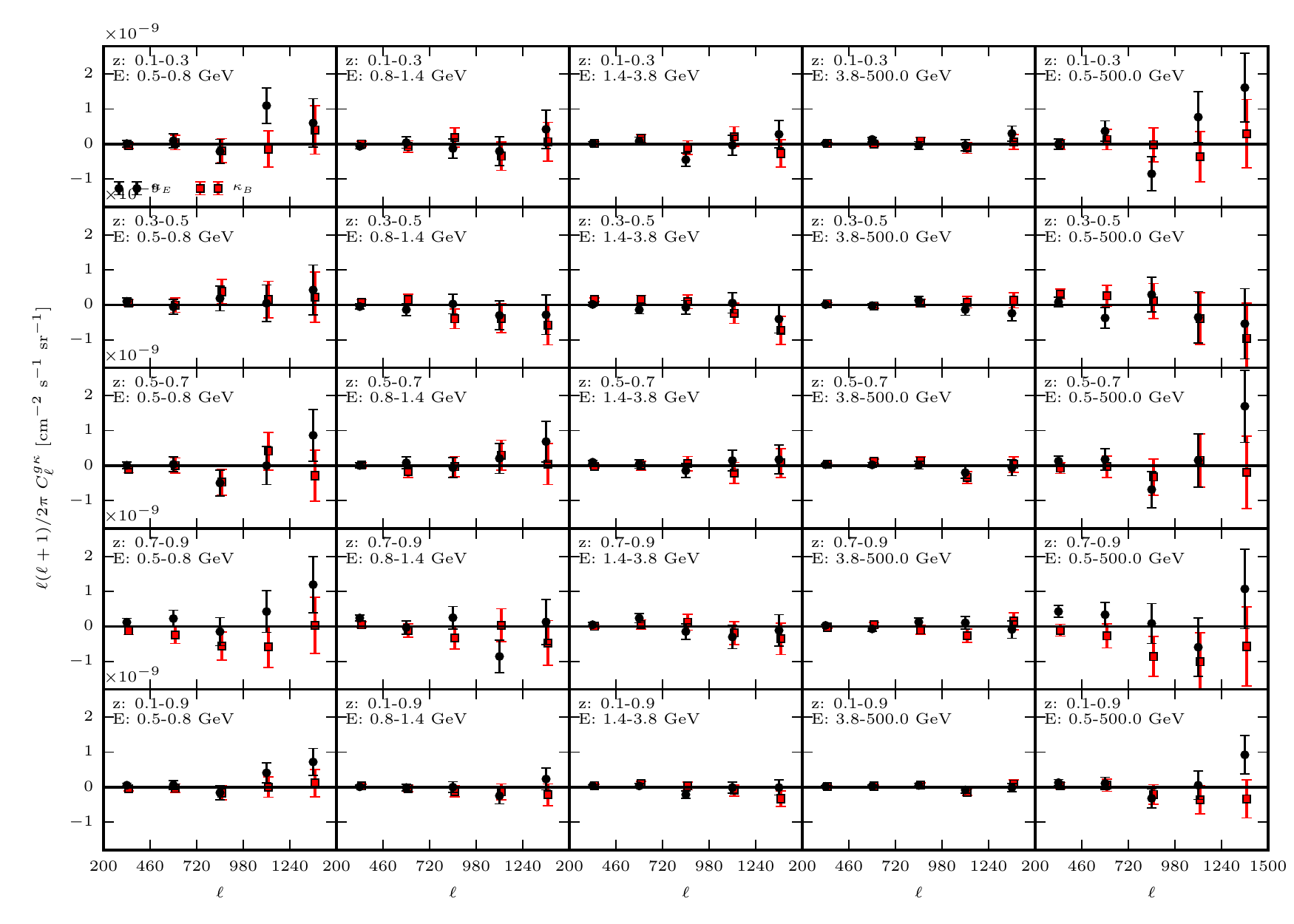}
    \caption{Measurement of the cross-spectrum $\hat C^{g\kappa}_\ell$ between \Fermi gamma rays and weak lensing data from KiDS for five energy bins for gamma-ray photons and five redshift bins for KiDS galaxies (black points). The cross-spectrum of the gamma rays and KiDS B-modes are depicted as red data points.}    
    \label{fig:KiDS-Cl-5x5}
    \end{center}
\end{figure*}

We present the measurement of the cross-correlation of \Fermi gamma rays with CFHTLenS, RCSLenS, and KiDS weak lensing data in Figs.~\ref{fig:CFHTLenS-Cl-5}, \ref{fig:RCSLenS-Cl-5}, and \ref{fig:KiDS-Cl-5x5}, respectively. The measurements for CFHTLenS and RCSLenS use a single redshift bin and the five energy bins described in Section~\ref{sec:data-Fermi}. The measurements for KiDS use the same energy bins but are further divided into the five redshift bins given in Section~\ref{sec:data-KiDS}.

Beside the cross-correlation of the gamma rays and shear due to gravitational lensing (denoted by black circles), we also show the cross-correlation between gamma rays and the B-mode of the shear as red squares. The B-mode of the shear is obtained by rotating the galaxy orientations by 45\degr, which destroys the gravitational lensing signal. Any significant B-mode signal would be indicative of spurious systematics in the lensing data.

\begin{figure}
    \begin{center}
	\includegraphics[width=\columnwidth]{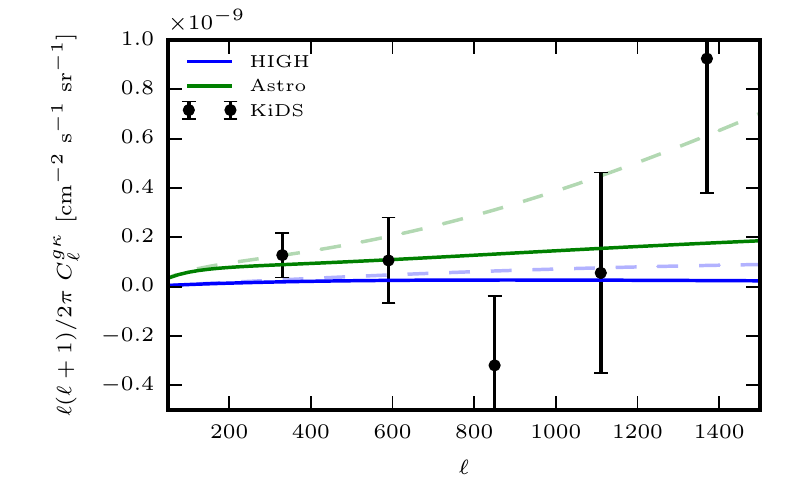}
  	\caption{Measurement of the cross-spectrum $\hat C^{g\kappa}_\ell$ between \Fermi gamma rays in the energy range 0.5--500 GeV and KiDS weak lensing data in the redshift range 0.1 -- 0.9 (black data points), compared to the expected signal from the sum of astrophysical sources (solid green) and annihilating \DM (solid blue). The astrophysical sources considered are blazars, mAGN, and SFG. The annihilating \DM model assumes the \high\ scenario, $\mdm=100$ GeV, and $\sv=3\ee{-26}\units{cm^3 s^{-1}}$. The dashed lines show the same models but without correcting for the \Fermi PSF.}
  	\label{fig:modelCldata}
   \end{center}
\end{figure}

The $\chi_0^2$ values of the measurements with respect to the hypothesis of a null signal, i.e., $\vec\mu=0$, are listed in Table~\ref{tab:chi2}. The $\chi_0^2$ values are consistent with a non-detection of a cross-correlation for all measurements. This finding is in agreement with the previous studies \cite{Shirasaki2014, Shirasaki2016} of cross-correlations between gamma rays and galaxy lensing. 
For a $3\ \sigma$ detection of the cross-correlation with astrophysical sources\footnote{Cross-correlations between tracers of LSS and gamma rays have already been detected in \cite{Xia2015}. A significant signal in the case of future weak lensing surveys is therefore a reasonable expectation.}, the error bars would have to shrink by a factor of 3 with respect to the current error bars for KiDS. This corresponds to a $\sim 4000$ deg$^2$ survey with KiDS characteristics, comparable in size to the galaxy surveys used in \cite{Xia2015}. 
This is further illustrated in Fig.~\ref{fig:modelCldata}, which shows the measurement for KiDS for the unbinned case in comparison with the expected correlation signal from astrophysical sources and annihilating \DM for the \high\ scenario and $\sv=3\ee{-26}\units{cm^3 s^{-1}}$ for $\mdm=100\units{GeV}$ and the $b\bar b$ channel. 
While these signals are not observable at current sensitivities, they are within reach of upcoming surveys, such as DES\footnote{https://www.darkenergysurvey.org/}.

The B-mode signal is consistent with zero for all measurements. We are thus confident that the measurement is not significantly contaminated by lensing systematics. 
At very small scales, lens-source clustering can cause a suppression of the lensing signal \citep{VanUitert2011,Hoekstra2015}. The angular scales we are probing in this analysis are however not affected by this.

\begin{table*}
    \begin{center}
         \caption{$\chi_0^2$ values with respect to the hypothesis of a null signal for the measurements of $\hat C_\ell^{g\kappa}$ shown in Figs.~\ref{fig:CFHTLenS-Cl-5}, \ref{fig:RCSLenS-Cl-5}, and~\ref{fig:KiDS-Cl-5x5}. The number of degrees of freedom is the number of multipole bins, i.e., $\nu=5$ for all measurements.}
	\begin{tabular}{lccccc}
\hline
&\multicolumn{5}{c}{$\chi_0^2\left(\hat C_\ell^{g\kappa}, \nu=5\right)$} \\
Energy bin [GeV]   &0.5 -- 0.8   &0.8 -- 1.4   &1.4 -- 3.8   &3.8 -- 500.0   &0.5 -- 500.0\\
\hline
CFHTLenS   &4.49   &7.77   &3.78   &8.43   &2.43\\
RCSLenS   &6.06   &6.75   &2.39   &6.47   &3.19\\
KiDS $0.1\leq z_\mathrm{B} < 0.3$   &5.96   &1.85   &6.53   &6.89   &8.47\\
KiDS $0.3\leq z_\mathrm{B} < 0.5$   &1.84   &1.94   &2.75   &3.42   &2.77\\
KiDS $0.5\leq z_\mathrm{B} < 0.7$   &3.27   &1.89   &4.02   &2.56   &5.57\\
KiDS $0.7\leq z_\mathrm{B} < 0.9$   &4.82   &11.42   &4.98   &2.88   &8.76\\
KiDS $0.1\leq z_\mathrm{B} < 0.9$   &7.16   &1.81   &5.42   &3.05   &6.55\\
\hline
	\end{tabular}
	\label{tab:chi2}
   \end{center}
\end{table*}%

\subsection{Interpretation}
\label{sec:DM-interpretation}
We wish to exploit the measurements presented in the previous subsection to derive constraints on WIMP \DM annihilation or decay. To derive the exclusion limits on the annihilation cross-section $\sv$ and WIMP mass $\mdm$, and the decay rate $\Gamma_\mathrm{dec}$ and $\mdm$, we apply the formalism described in Section~\ref{sec:stats}.

\begin{figure}
    \includegraphics[width=1.0\columnwidth]{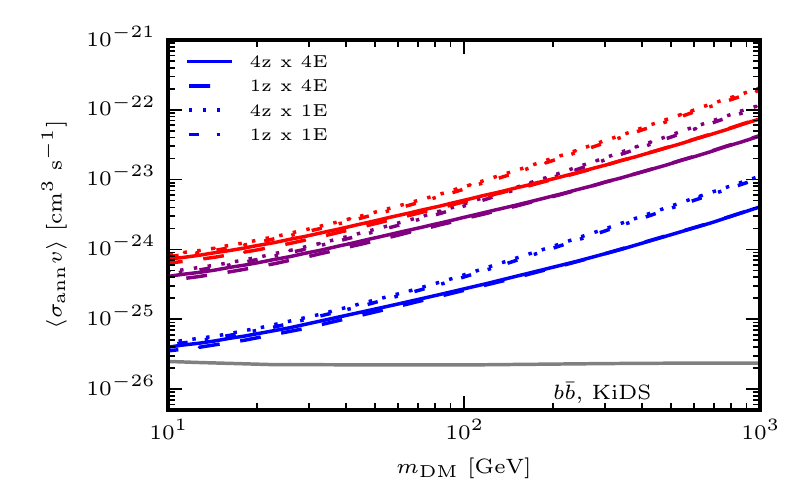}
    \caption{Exclusion limits on the annihilation cross-section $\sv$ and WIMP mass $\mdm$ for the clustering scenarios \high\ (blue), \ave\ (purple), and \low\ (red) and for different binning strategies for the KiDS data. The lines correspond to $2\sigma$ upper limits on $\sv$ and $\mdm$, assuming a 100\% branching ratio into $b\bar b$. No binning in redshift or energy (1z x 1E) is denoted by dash-dotted lines. The case of binning in redshift but not energy (4z x 1E) is plotted as dotted lines, while binning in energy but not redshift (1z x 4E) is plotted as dashed lines. Finally, binning in both redshift and energy (4z x 4E) is shown as solid lines. The thermal relic cross-section, from \protect\cite{Steigman2012}, is shown in grey.}
    \label{fig:limits-binning}
\end{figure}

\begin{figure}
    \includegraphics[width=1.0\columnwidth]{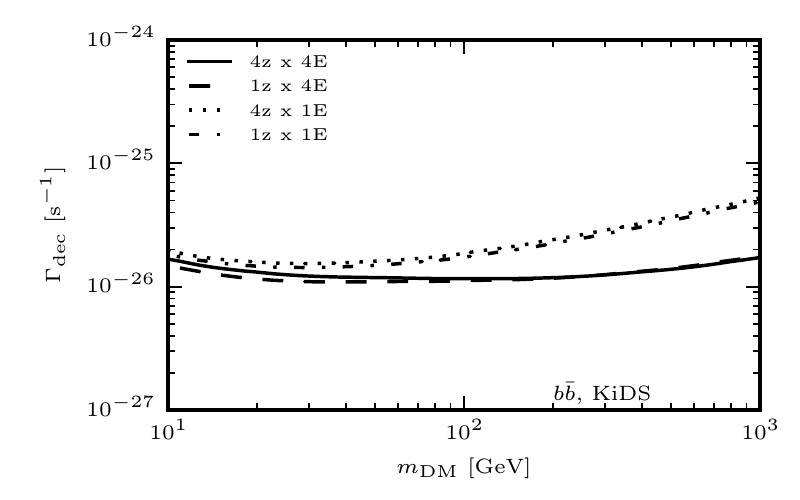}
    \caption{Exclusion limits on the decay rate $\Gamma_\mathrm{dec}$ and WIMP mass $\mdm$ for the $b\bar b$ channel for different binning strategies for the KiDS data. The style of the lines is analogous to Fig.~\ref{fig:limits-binning}.}
    \label{fig:dDM-limits-binning}
\end{figure}

In \cite{Camera2015} it was shown that the spectral and tomographic information contained within the gamma-ray and lensing data can improve the limits on $\sv$ and $\Gamma_\mr{dec}$. We show the effect of different combinations of spectral and tomographic binning for the case of KiDS and annihilations into $b\bar b$ pairs under the \high\ scenario in Fig.~\ref{fig:limits-binning} and for \DM decay in Fig.~\ref{fig:dDM-limits-binning}. For these limits we adopt the conservative assumption that all gamma rays are sourced by \DM, i.e., no astrophysical contributions are included. There is a significant improvement of the limits when using four energy bins over a single energy bin, especially at high particle masses $\mdm$. 
This is due to the fact that the UGRB scales roughly as $E^{-2.3}$ \citep{Ackermann2015}. The vast majority of the photons in the 0.5 -- 500 GeV bin therefore come from low energies. However, the peak in the prompt gamma-ray emission induced by \DM occurs at energy $\sim \mdm/20$ (annihilating) or $\sim \mdm/40$ (decaying) for $b\bar b$ and at higher energies for the other channels. Thus, for high $\mdm$, a single energy bin of 0.5 -- 500 GeV largely increases the noise without significantly increasing the expected \DM signal with respect to the 3.8 -- 500 GeV bin. 

The improvement due to tomographic binning is only marginal. Two factors contribute to this lack of improvement: Firstly, in the case of no observed correlation signal -- as is the case here -- the differences in the redshift dependence of the astrophysical and \DM sources do not come to bear because there is no signal to disentangle. Secondly, the lensing window functions are quite broad and thus insensitive to the featureless window function of the \DM gamma-ray emissions, as depicted in Fig.~\ref{fig:windowfunctions}. This is due to the cumulative nature of lensing on the one hand and the fact that photo-z's cause the true $n(z)$ to be broader than the redshift cuts we impose on the other hand.
This is in contrast with spectral binning, which allows us to sharply probe the characteristic gamma-ray spectrum induced by \DM. As shown in Fig.~\ref{fig:spectra}, annihilating \DM shows a pronounced pion bump when annihilating into $b\bar b$ and a cutoff corresponding to the \DM mass $\mdm$, while for decaying \DM the cutoff appears at half the \DM mass. 
For this reason we refrain from a tomographic analysis for CFHTLenS and RCSLenS as we expect little to no improvements of the limits.

The limits can be further tightened by taking into account known astrophysical sources of gamma rays. This comes, however, at the expense of introducing new uncertainties in the modelling of said astrophysical sources. Going forward, we include the astrophysical sources to show the sensitivity reach of such analyses but also show the conservative limits derived under the assumption that all gamma rays are sourced by \DM.

To account for the astrophysical sources, we subtract the combination of the three populations (blazars, mAGN, and SFG) described in Section~\ref{sec:models} from the observed cross-correlation signal. The \DM limits are then obtained by proceeding as before but using the residuals between the cross-correlation measurement and the astrophysical contribution. Since we assume no error on the astrophysical models, the limits obtained by including blazars, mAGN, and SFG contributions should be considered as a sensitivity reach for a future situation where gamma-ray emission from these astrophysical sources will be perfectly understood.

\begin{figure*}
    \includegraphics[width=1.0\textwidth]{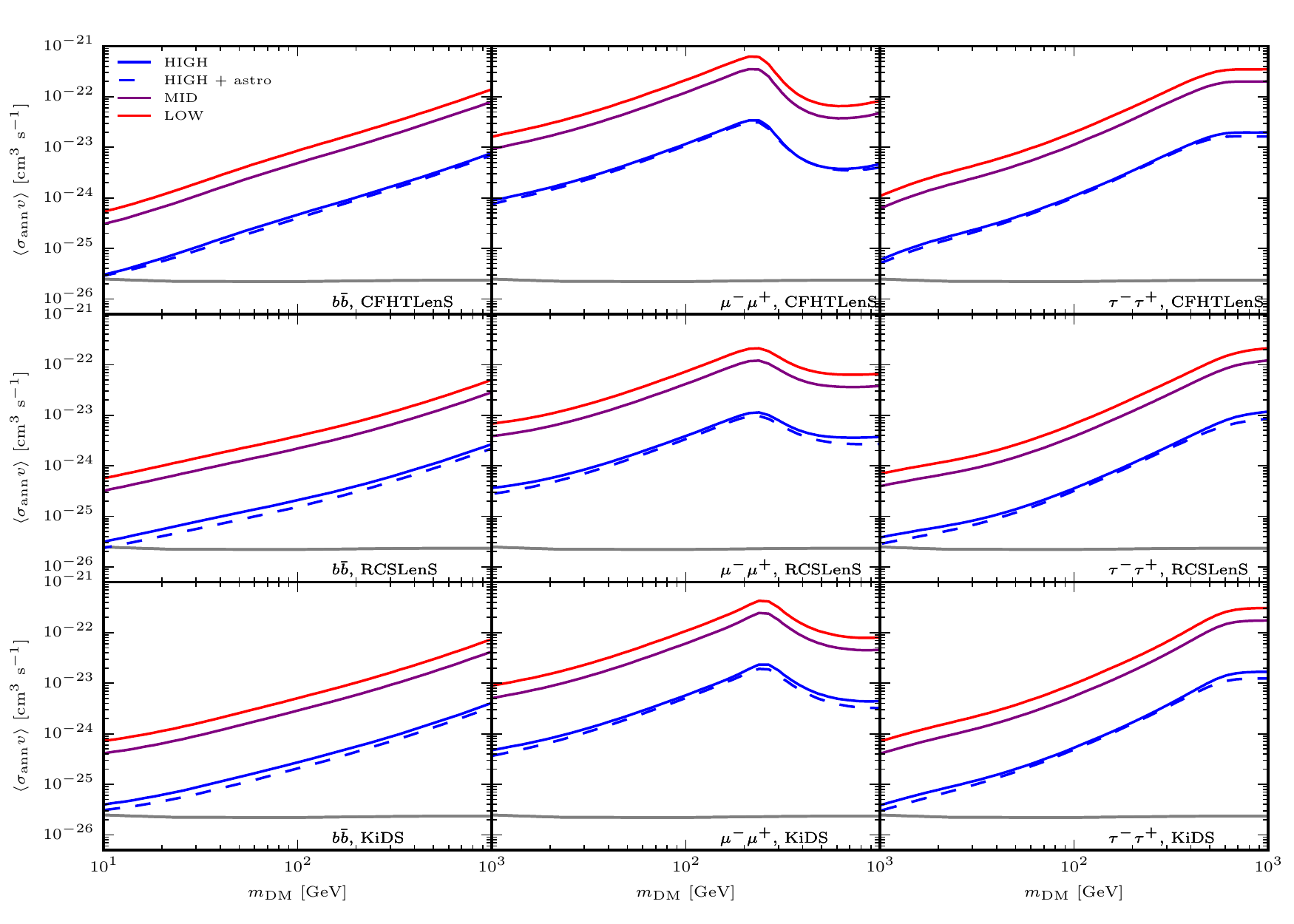}
    \caption{Exclusion limits on the annihilation cross-section $\sv$ and WIMP mass $\mdm$ at $2\sigma$ significance for CFHTLenS, RCSLenS, and KiDS and annihilation channels $b\bar b$, $\mu^-\mu^+$, and $\tau^-\tau^+$. CFHTLenS and RCSLenS use four energy bins while KiDS additionally makes use of four redshift bins. The exclusion limits are for the three clustering scenarios \high\ (blue), \ave\ (purple), and \low\ (red). The dashed blue line indicates the improvement of the limits for the \high\ scenario when including the astrophysical sources in the analysis.}
    \label{fig:limits-separate}
\end{figure*}

\begin{figure*}
    \includegraphics[width=1.0\textwidth]{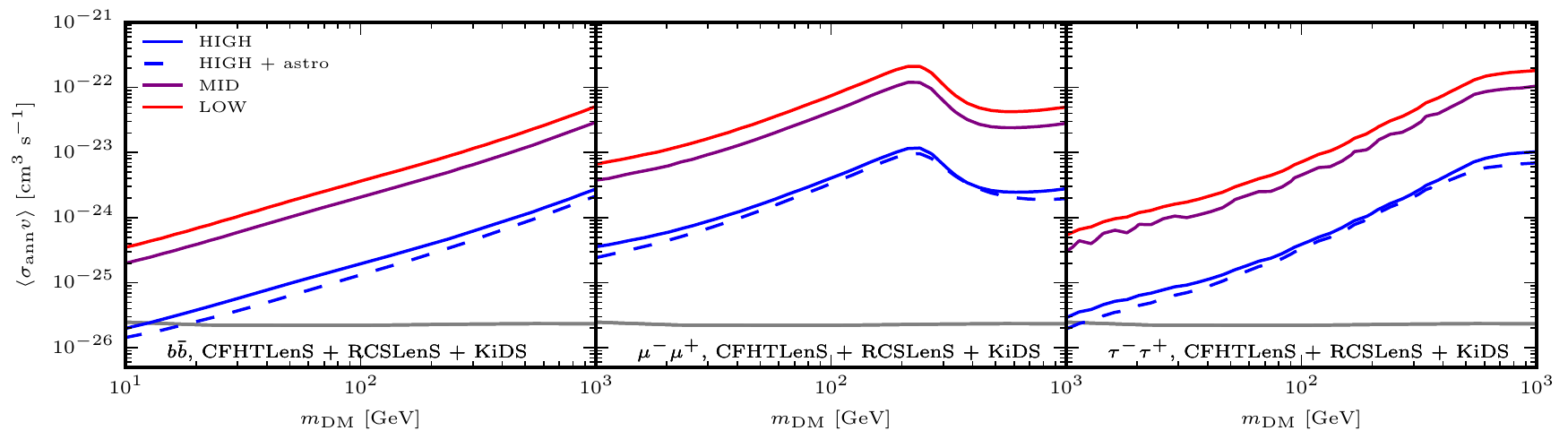}
    \caption{Exclusion limits on the annihilation cross-section $\sv$ and WIMP mass $\mdm$ at $2\sigma$ significance for the combination of CFHTLenS, RCSLenS, and KiDS. The style of the lines is the same as for Fig.~\ref{fig:limits-separate}.}
    \label{fig:limits-combined}
\end{figure*}

\begin{figure*}
    \includegraphics[width=1.0\textwidth]{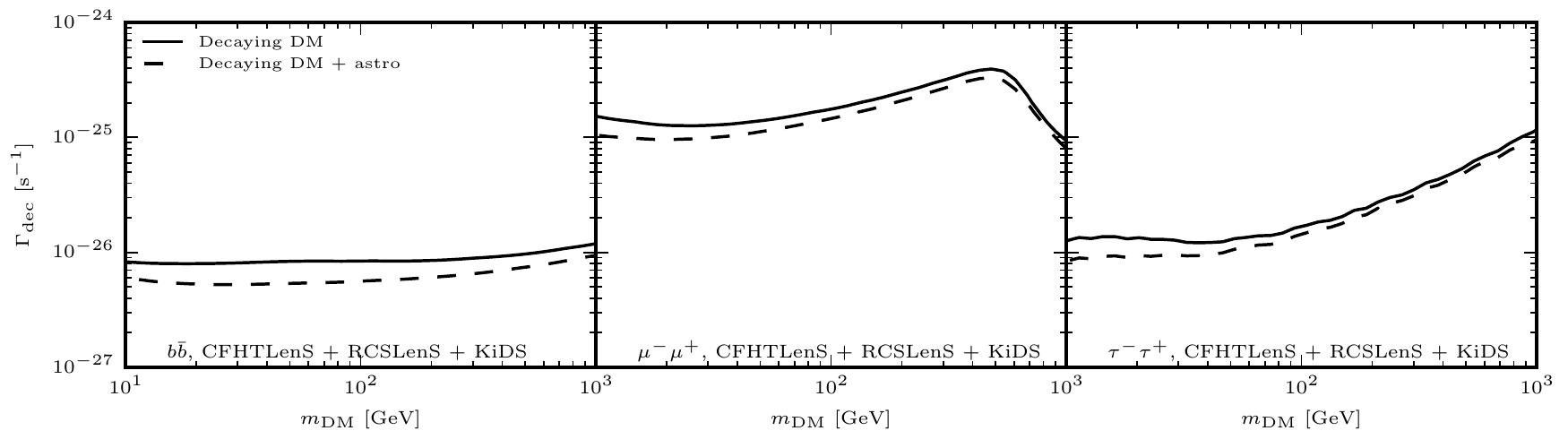}
    \caption{Exclusion limits on the decay rate $\Gamma_\mathrm{dec}$ and WIMP mass $\mdm$ at $2\sigma$ significance for the combination of CFHTLenS, RCSLenS, and KiDS (solid black). Including the astrophysical sources in the analysis results in the more stringent exclusion limits denoted by the black dashed line.}
    \label{fig:dDM-limits-combined}
\end{figure*}

The resulting $2\sigma$ exclusion limits on the \DM annihilation cross-section $\sv$ for the $b\bar b$, $\mu^-\mu^+$, and $\tau^-\tau^+$ channels are shown in Fig.~\ref{fig:limits-separate}. Finally, the combined exclusion limits for CFHTLenS, RCSLenS, and KiDS are shown in Fig.~\ref{fig:limits-combined} and Fig.~\ref{fig:dDM-limits-combined} for annihilating and decaying WIMP \DM, respectively. The exclusion limits for annihilating \DM should be compared to the thermal relic cross-section \citep{Steigman2012}, shown in grey. Under optimistic assumptions about the clustering of \DM, i.e., the \high\ model, and accounting for contributions from astrophysical sources (dashed blue line), we can exclude the thermal relic cross-section for masses $\mdm\la 20$ GeV for the $b\bar b$ channel. 
In the case of annihilations or decays into muons or tau leptons, the exclusion limits change shape and become stronger for large \DM masses, compared to the $b$ channel. This is due to the fact that, for heavy \DM candidates, inverse Compton scattering produces a significant amount of gamma-ray emission in the upper energy range probed by our measurement \citep{Ando2016}.
If we make the conservative assumption that only \DM contributes to the UGRB, i.e., we do not account for the astrophysical sources of gamma rays, the exclusion limits weaken slightly, as seen in the difference between the dashed and solid blue lines in Fig.~\ref{fig:limits-combined}. In this case the thermal relic cross-section can be excluded for $\mdm\la 10$ GeV for the $b\bar b$ channel. These limits are consistent with those forecasted in \cite{Camera2015}.

The exclusion limits when \DM is assumed to be the only contributor to the UGRB are comparable to those derived from the energy spectrum of the UGRB in \cite{The-Fermi-LAT-Collaboration2015a}. However, when the contribution from astrophysical sources is accounted for, the limits in \cite{The-Fermi-LAT-Collaboration2015a} improve by approximately one order of magnitude, while our limits see only modest improvements. This is due to the fact that we do not observe a cross-correlation signal. The constraining power therefore largely depends on the size of the error bars. The contribution from astrophysical sources is small compared to the size of our error bars, as shown in Fig.~\ref{fig:modelCldata}, explaining the modest gain in constraining power when including the astrophysical sources compared to probes that observe a signal.
The exclusion limits obtained in \cite{Fornasa2016} from the measurement of the UGRB angular auto-power spectrum are stronger than the ones presented here. Those limits are dominated by the emission from \DM subhaloes in the Milky Way, a component that is not considered in our analysis since it does not correlate with  weak lensing. When restricting the analysis of the auto-spectrum in \cite{Fornasa2016} to only the extragalactic components, our cross-correlation analysis yields more stringent limits. The limits presented here are comparable to those of similar analyses of the cross-correlation between gamma rays and weak lensing \citep{Shirasaki2014, Shirasaki2016} but weaker than those derived from cross-correlations between gamma rays and galaxy surveys \citep{Cuoco2015,Regis2015}.
The exclusion limits from all these extragalactic probes are somewhat weaker than those derived from dSphs \citep{Ackermann2015b,Baring2016}.

The weaker limits obtained when using KiDS data, compared to those obtained from RCSLenS data, can be traced to the high data point at small scales in the low energy bins.
Restricting the analysis to the $\ell$-range of 200 to 1240, i.e., removing the last data point, improves the limits derived from the KiDS to exceed those derived from RCSLenS, as one would expect from the covariances of the two measurements.
To check whether the high data point is part of a trend that might become significant at even smaller scales, we extend the measurement to higher $\ell$-modes. Doing so reveals a high scatter of the data points around zero beyond $\ell\ga 1500$, and no further excess of power at smaller scales. It should be noted that at these small scales, we are probing close to the pixel scale and are within the \Fermi PSF, so the signal is expected to be consistent with zero there.
Including astrophysical sources absorbs some of the effect of the high data point at small scales. The limits including astrophysical sources of gamma rays are thus closer than those assuming only \DM as the source of gamma rays.

\section{Conclusion}
\label{sec:conclusions}
We have measured the angular cross-power spectrum of \Fermi gamma rays and weak gravitational lensing data from CFHTLenS, RCSLenS, and KiDS. Combined together, the three surveys span a total area of more than 1000 deg$^2$. We made use of 8 years of Pass 8 \Fermi data in the energy range 0.5 -- 500 GeV which was divided further into four energy bins. 
For CFHTLenS and RCSLenS, the measurement was done for a single redshift bin, while the KiDS data were further split into five redshift bins, making this the first measurement of tomographic weak lensing cross-correlation.
We find no evidence of a cross-correlation signal in the multipole range $200 \leq \ell < 1500$, consistent with previous studies and forecasts based on the expected signal and current error bars.

Using these measurements we constrain the WIMP \DM annihilation cross-section $\sv$ and decay rate $\Gamma_\mr{dec}$ for WIMP masses between 10 GeV and 1 TeV. 
Assuming the \high\ model for small-scale clustering of \DM and accounting for astrophysical sources, we are able to exclude the thermal annihilation cross-section for WIMPs of masses up to $\sim 20$ GeV for the $b\bar b$ channel. 
Not accounting for the astrophysical contribution weakens the limits only slightly, while the exclusion limits for the more conservative clustering models \ave\ and \low\ are a factor of $\sim10$ weaker. We find that tomography does not significantly improve the constraints. However, exploiting the spectral information of the gamma rays strengthens the limits by up to a factor $\sim 3$ at high masses.

The exclusion limits derived in this work are competitive with others derived from the UGBR, such as its intensity energy spectrum \citep{The-Fermi-LAT-Collaboration2015a}, auto-power spectrum \citep{Fornasa2016}, cross-correlation with weak lensing \mbox{\citep{Shirasaki2014, Shirasaki2016}} or galaxy surveys \citep{Regis2015, Cuoco2015}. Exclusion limits derived from local probes, such as dSphs, are stronger, however \citep{Ackermann2015b}.

Future avenues to build upon this analysis include the use of upcoming large area lensing data sets, such as future KiDS data, DES, HSC\footnote{http://www.naoj.org/Projects/HSC/}, LSST\footnote{http://www.lsst.org/}, and Euclid\footnote{http://sci.esa.int/euclid/}, which will make it possible to detect a cross-correlation signal between gamma rays and gravitational lensing. The analysis would also benefit from extending the range of the gamma-ray energies covered, by making use of measurements from atmospheric Cherenkov telescopes, which are more sensitive to high-energy photons \citep{Ripken2014}. 

Instead of treating the astrophysical contributions as a contamination to a \DM signal, the measurements presented in this work could be used to investigate the astrophysical extragalactic gamma-ray populations that are thought to be responsible for the UGRB. We defer this to a future analysis.

\section*{Acknowledgements}
This work is based on observations obtained with MegaPrime/MegaCam, a joint project of CFHT and CEA/IRFU, at the Canada-France-Hawaii Telescope (CFHT) which is operated by the National Research Council (NRC) of Canada, the Institut National des Sciences de l'Univers of the Centre National de la Recherche Scientifique (CNRS) of France, and the University of Hawaii. This research used the facilities of the Canadian Astronomy Data Centre operated by the National Research Council of Canada with the support of the Canadian Space Agency. CFHTLenS and RCSLenS data processing was made possible thanks to significant computing support from the NSERC Research Tools and Instruments grant program.

This work is based on data products from observations made with ESO Telescopes at the La Silla Paranal Observatory under programme IDs 177.A-3016, 177.A-3017 and 177.A-3018, and on data products produced by Target/OmegaCEN, INAF-OACN, INAF-OAPD and the KiDS production team, on behalf of the KiDS consortium. OmegaCEN and the KiDS production team acknowledge support by NOVA and NWO-M grants. Members of INAF-OAPD and INAF-OACN also acknowledge the support from the Department of Physics \& Astronomy of the University of Padova, and of the Department of Physics of Univ. Federico II (Naples).

Computations for the $N$-body simulations were performed in part on the Orcinus supercomputer at the WestGrid HPC consortium (www.westgrid.ca), in part on the GPC supercomputer at the SciNet HPC Consortium. SciNet is funded by: the Canada Foundation for Innovation under the auspices of Compute Canada; the Government of Ontario; Ontario Research Fund -- Research Excellence; and the University of Toronto. 

TT and LvW are funded by NSERC and CIfAR. TT is furthermore supported by the Swiss National Science Foundation. 
SC is supported by ERC Starting Grant No. 280127.
MF and SA acknowledge support from Netherlands Organization for Scientific Research (NWO) through a Vidi grant.
MR and NF are supported by the research grant Theoretical Astroparticle Physics number 2012CPPYP7 under the program PRIN 2012 funded by the Ministero dell'Istruzione, Universit\`a  e della Ricerca (MIUR); by the research grants TAsP (Theoretical Astroparticle Physics) and Fermi funded by the Istituto Nazionale di Fisica Nucleare (INFN); by the Excellent Young PI Grant: The Particle Dark-matter Quest in the Extra-galactic Sky.
JHD acknowledges support from the European Commission under a Marie-Sk{l}odwoska-Curie European Fellowship (EU project 656869).
MB is supported by the Netherlands Organization for Scientific Research, NWO, through grant number 614.001.451, and by the European Research Council through FP7 grant number 279396.
TE is supported by the Deutsche Forschungsgemeinschaft in the framework of the TR33 `The Dark Universe'.
CH acknowledges support from the European Research Council under grant number 647112.
HHi is supported by an Emmy Noether grant (No. Hi 1495/2-1) of the Deutsche Forschungsgemeinschaft.
HHo acknowledges support from the European Research Council under FP7 grant number 279396.
KK acknowledges support by the Alexander von Humboldt Foundation.
MV acknowledges support from the European Research Council under FP7 grant number 279396 and the Netherlands Organisation for Scientific Research (NWO) through grants 614.001.103.

{\small \textit{Author Contributions:} All authors contributed to the development and writing of this paper. The authorship list is given in three groups: the lead authors (TT, SC, MF, MR, LvW), followed by two alphabetical groups. The first alphabetical group includes those who are key contributors to both the scientific analysis and the data products. The second group covers those who have either made a significant contribution to the data products, or to the scientific analysis.}

\appendix
\section{Fourier-space estimator performance}
\label{sec:Cltest}
To check the ability of the power spectrum estimator given by Eq.~\eqref{equ:xi2Cl} to recover the true underlying power spectrum we test it on a suite of mock simulations. Specifically, we compare the auto-spectrum of the convergence with the estimated cross-spectrum between the convergence and the shear, which are expected to yield the same result. This is analogous to the cross-spectrum of the gamma rays and shear but easier to handle, as high-resolution simulation products for convergence and shear are readily available. 

The simulation products we use are part of the Scinet LIght Cone Simulation suite \cite[][SLICS hereafter]{Harnois-Deraps2015a}, which consist of 930 realizations of lensing data over $10\times10$deg$^2$ patches in a {\it WMAP}9+SN+BAO cosmology ($\{\Omega_{\rm M}, \Omega_{\Lambda},  \Omega_{b}, \sigma_8, h, n_s \} = \{0.2905, 0.7095, 0.0473, 0.826, 0.6898, 0.969 \}$). 
The convergence and two shear components are constructed by ray-tracing up to 18 density planes between redshift zero and 3, and finally mapped on to 7745$^2$ pixels (see the SLICS reference for details about how this is implemented numerically).
For our particular setup, we use the maps constructed while assuming that the galaxy sources 
are all placed at redshift 0.582. This is of course not representative of the real galaxy distribution of the data, 
but closely matches the mean of the distribution, which is sufficient for the purpose of calibration.

For the purpose of the verification of our estimator in Eq.~\eqref{equ:xi2Cl}, we use a subset of 100 realizations. The convergence and shear maps are cropped to 7700$^2$ pixels and then down-sampled by a factor of 10 to closer resemble the pixel size encountered in the gamma-ray analysis.

\begin{figure}
    \begin{center}
    \includegraphics[width=\columnwidth]{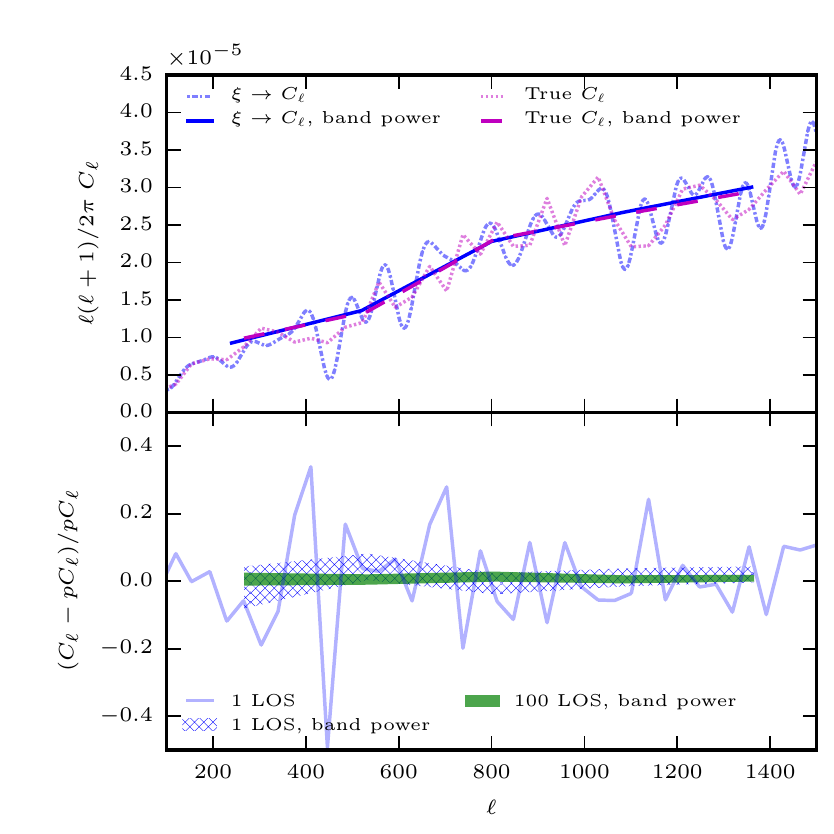}
    \caption{\emph{Top}: power spectrum estimated using Eq.~\eqref{equ:xi2Cl} (dash-dotted blue), band power (solid blue), true power spectrum (dotted magenta), and true band power (dashed magenta) for one line-of-sight. \emph{Bottom}: difference between estimated and true power spectrum (light solid blue) for one line-of-sight, difference between the estimated and true band power for one line-of-sight (hashed blue) and 100 line-of-sights (solid green).}
    \label{fig:Cltest}
    \end{center}
\end{figure}

We measure the tangential shear correlation function between the convergence and shear maps using the same binning scheme as the gamma-ray cross-correlation measurement, i.e., 300 linearly spaced bins between 1 and 301 arcmin. The power spectrum estimated using Eq.~\eqref{equ:xi2Cl} is then expected to agree with the auto-power spectrum of the convergence map $C_\ell^{\kappa\kappa}$. The power spectra measured on the simulations are shown in Fig.~\ref{fig:Cltest}. For the scales of interest in this work the estimator recovers the power spectrum to within 5 per cent on individual line-of-sights, which is within the error on the mean per $\ell$-bin of the true power spectrum. The agreement is within $\sim 1$ per cent for 100 line-of-sights, which is within the error on the mean of the 100 true power spectra, showing that the fluctuations seen on individual line-of-sight average out.

\begin{figure}
    \begin{center}
    \includegraphics[width=\columnwidth]{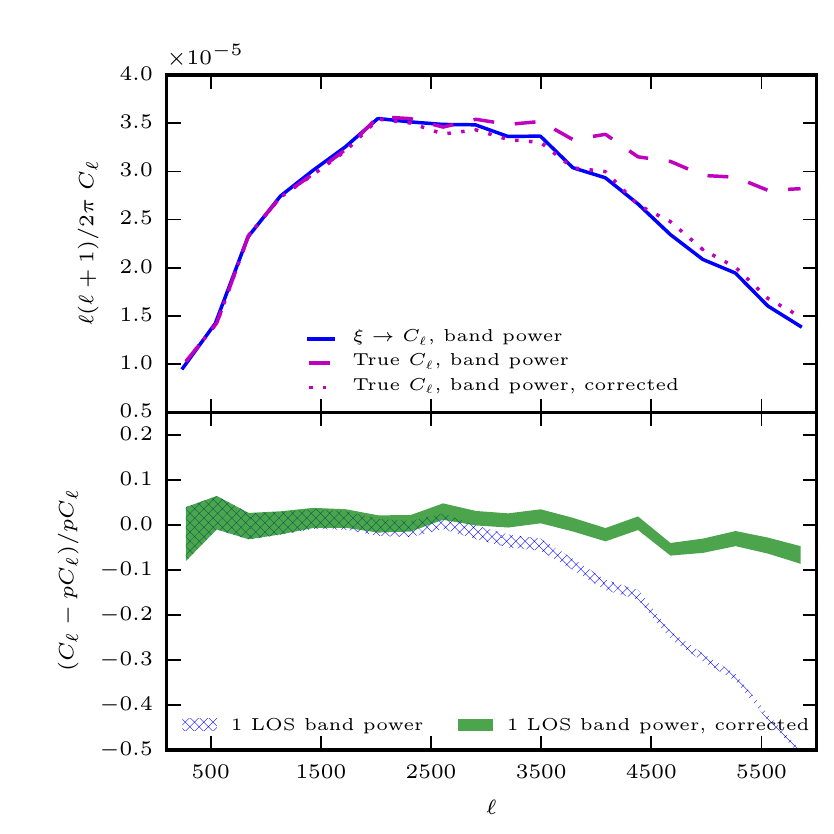}
    \caption{\emph{Top}: band power spectrum estimated using Eq.~\eqref{equ:xi2Cl} (solid blue), true band power (dashed magenta), and true band power corrected for the effect of finite $\vartheta_\mr{min}$ in Eq.~\eqref{equ:xi2Cl-thetamin}(dotted magenta). \emph{Bottom}: difference between estimated band power and true band power (solid green) and corrected for finite $\vartheta_\mr{min}$ (hashed blue).}
    \label{fig:Cltest-highres}
    \end{center}
\end{figure}

One caveat is that the range of integration in Eq.~\eqref{equ:xi2Cl} is formally from 0 to infinite angular separation. By restricting the integration to some finite range $\vartheta_\mr{min}$ to $\vartheta_\mr{max}$, the resulting power spectrum estimate can become biased. 

To estimate the effect of restricting the angular range of the two-point correlation function on the power spectrum estimate we produce a high-resolution measurement of the convergence power spectrum $C_\ell^{\kappa\kappa}$. Using the relation in Eq.~\eqref{equ:Cl2xi} between the power spectrum $C_\ell^{\kappa\kappa}$ and the tangential shear correlation function $\xi^{\kappa\gamma_t}(\vartheta)$, we compute a theory estimate $\xi^{\kappa\gamma_t}_\mr{th}(\vartheta)$ from the measured high-resolution convergence power spectrum. Alternatively, we could also have used an analytical model for the power spectrum or correlation function. However, in order to make the comparison between the different power spectrum estimates as direct as possible, we chose to minimize the amount of external information.

Finally, to estimate the effect of restricting the range of integration we compute the corrections terms
\begin{splitequation}
	\label{equ:xi2Cl-thetamin}
	C_\ell^{\vartheta_\mr{min}}  = 2\pi \int_0^{\vartheta_\mr{min}} \diff \vartheta'\ \vartheta J_2(\ell\vartheta') \xi^{\kappa\gamma_t}_\mr{th}(\vartheta')
\end{splitequation}
and
\begin{splitequation}
	C_\ell^{\vartheta_\mr{max}}  = 2\pi \int_{\vartheta_\mr{max}}^\infty \diff \vartheta'\ \vartheta J_2(\ell\vartheta') \xi^{\kappa\gamma_t}_\mr{th}(\vartheta') \ .
\end{splitequation}
The effect of the minimum angular separation $\vartheta_\mr{min}$ is a suppression of power at small scales, restricting the range of scales where the power spectrum estimate is unbiased. However, by forward modelling, i.e., accounting for the effect of the minimum angular separation when comparing the measurements to models, the effective range can be increased significantly. This can be seen in Fig.~\ref{fig:Cltest-highres}, where accounting for $C_\ell^{\vartheta_\mr{min}}$ increases the range of validity from $\ell \sim 2000$ to $\ell \sim 6000$. On the scales considered in the main body of this work ($\ell \leq 1500$) the effect of a finite $\vartheta$ is negligible and no forward modelling of this effect was conducted.

A finite maximum angular separation $\vartheta_\mr{max}$ does not lead to a systematic bias in the power spectrum estimation like the effect of $\vartheta_\mr{min}$. The oscillatory nature of the estimated power spectrum requires the use of band power, however. The width of the bins results in an effective lower limit on the scales that can be estimated. For fixed $\vartheta'$, the Bessel function in Eq.~\eqref{equ:xi2Cl} oscillates with period of $\sim\frac{2\pi}{\vartheta'}$. The shortest period that can be probed is therefore $\sim\frac{2\pi}{\vartheta'_\mr{max}}$. Requiring two to three periods per $\ell$-bin, the minimum reliable bin width for a maximum angular separation of $\vartheta_\mr{max} = 301\units{arcmin}$ is therefore $\Delta_\ell \sim 200$. The bin width chosen in this work is $\Delta_\ell = 260$ and can thus be assumed to yield a reliable estimate of the power spectrum.

\begin{figure}
    \begin{center}
    \includegraphics[width=\columnwidth]{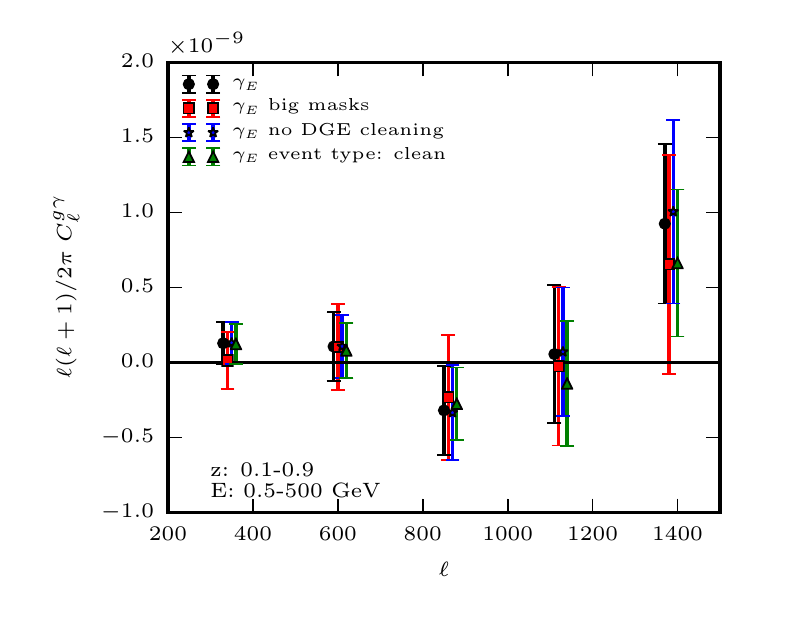}
    \caption{Measurement of the cross-spectrum $\hat C^{g\kappa}_\ell$ between \Fermi gamma rays in the energy range 0.5--500 GeV and KiDS weak lensing data in the redshift range 0.1 -- 0.9 for different gamma-ray data preparation choices: fiducial, as described in Section~\ref{sec:data-Fermi} (black points); using two-degree radius circular masks for all point sources (red squares); no cleaning of the diffuse Galactic emission (DGE) (blue stars); and using the \texttt{clean} event selection (green triangles).}
    \label{fig:dataselection}
    \end{center}
\end{figure}

\bibliographystyle{mnras}
\bibliography{gamma-lensing2.bib}

\bsp	
\label{lastpage} 
\end{document}